\documentclass[onecolappendix]{aastex63}
\usepackage{amsmath}
\usepackage{amsfonts}
\usepackage{amssymb}
\usepackage{xcolor}
\usepackage{longtable}
\usepackage{makecell}
\usepackage{graphicx}
\usepackage{hyperref}
\usepackage[T1]{fontenc}

\usepackage{comment}
\usepackage[encapsulated]{CJK}

\usepackage[utf8]{inputenc}

\newcommand{\cntext}[1]{\begin{CJK}{UTF8}{gbsn}#1\end{CJK}}

\newcommand{\rhigh}{R_{\text{high}}}
\newcommand{\rlow}{R_{\text{low}}}
\newcommand{\spin}{a_{*}}
\newcommand{\asym}{a_{1}}
\newcommand{\ipole}{{\tt ipole }}
\newcommand{\iharm}{{\tt iharm }}
\newcommand{\kharma}{{KHARMA }}
\newcommand{\patoka}{{\tt PATOKA }}

\defcitealias{EHTC_2019_1}{EHTC M87~I}
\defcitealias{EHTC_2019_2}{EHTC M87~II}
\defcitealias{EHTC_2019_3}{EHTC M87~III}
\defcitealias{EHTC_2019_4}{EHTC M87~IV}
\defcitealias{EHTC_2019_5}{EHTC M87~V}
\defcitealias{EHTC_2019_6}{EHTC M87~VI}
\defcitealias{EHTC_2021_7}{EHTC M87~VII}
\defcitealias{EHTC_2021_8}{EHTC M87~VIII}
\defcitealias{EHTC_2023_9}{EHTC M87~IX}

\defcitealias{EHTC_2022_1}{EHTC SgrA~I}
\defcitealias{EHTC_2022_2}{EHTC SgrA~II}
\defcitealias{EHTC_2022_3}{EHTC SgrA~III}
\defcitealias{EHTC_2022_4}{EHTC SgrA~IV}
\defcitealias{EHTC_2022_5}{EHTC SgrA~V}
\defcitealias{EHTC_2022_6}{EHTC SgrA~VI}

\defcitealias{EHTC_2024_1}{EHTC M87~2018~I}
\defcitealias{EHTC_2024_2}{EHTC SgrA~VII}
\defcitealias{EHTC_2024_3}{EHTC SgrA~VIII}
\defcitealias{EHTC_2025_1}{EHTC M87~2018~II}
\defcitealias{EHTC_2025_1_M87_21}{EHTC M87~2021}

\begin{document}

\title{Ring Asymmetry and Spin in M87*}

\correspondingauthor{Nicholas S. Conroy}\email{nconroy2@illinois.edu}

\author[0009-0000-1376-2352]{Vadim Bernshteyn} 
\affiliation{Department of Physics, University of Illinois, 1110 West Green Street, Urbana, IL 61801, USA}
\affiliation{Steward Observatory and Department of Astronomy, University of Arizona, 933 N. Cherry Ave., Tucson, AZ 85721, USA}

\author[0000-0003-2886-2377]{Nicholas S. Conroy}
\affiliation{Department of Astronomy, University of Illinois, 1002 West Green Street, Urbana, IL 61801, USA}

\author[0000-0002-5518-2812]{Michi Bauböck}
\affiliation{Department of Physics, University of Illinois, 1110 West Green Street, Urbana, IL 61801, USA}

\author[0000-0003-3826-5648]{Paul Tiede}
\affiliation{Center for Astrophysics $|$ Harvard \& Smithsonian, 60 Garden Street, Cambridge, MA 02138, USA}
\affiliation{Black Hole Initiative at Harvard University, 20 Garden Street, Cambridge, MA 02138, USA}

\author[0000-0002-2514-5965]{Abhishek~V.~Joshi}
\affil{Department of Physics, University of Illinois, 1110 West Green Street, Urbana, IL 61801, USA}
\affil{Illinois Center for Advanced Study of the Universe, 1110 West Green Street, Urbana, IL 61801, USA}

\author[0000-0002-0393-7734]{Ben~S.~Prather}
\affiliation{Department of Physics, University of Illinois, 1110 West Green Street, Urbana, IL 61801, USA}
\affiliation{Los Alamos National Lab, Los Alamos, NM, 87545, USA}

\author[0000-0001-7451-8935]{Charles~F.~Gammie}
\affiliation{Department of Astronomy, University of Illinois, 1002 West Green Street, Urbana, IL 61801, USA}
\affiliation{Department of Physics, University of Illinois, 1110 West Green Street, Urbana, IL 61801, USA}
\affiliation{Illinois Center for Advanced Study of the Universe, 1110 West Green Street, Urbana, IL 61801, USA}
\affiliation{National Center for Supercomputing Applications, 605 East Springfield Avenue, Champaign, IL 61820, USA}


\author{the Event Horizon Telescope Collaboration} 
\noaffiliation

\author[0000-0002-9475-4254]{Kazunori Akiyama}
\affiliation{Massachusetts Institute of Technology Haystack Observatory, 99 Millstone Road, Westford, MA 01886, USA}
\affiliation{National Astronomical Observatory of Japan, 2-21-1 Osawa, Mitaka, Tokyo 181-8588, Japan}
\affiliation{Black Hole Initiative at Harvard University, 20 Garden Street, Cambridge, MA 02138, USA}

\author[0000-0002-7816-6401]{Ezequiel Albentosa-Ruíz}
\affiliation{Departament d'Astronomia i Astrofísica, Universitat de València, C. Dr. Moliner 50, E-46100 Burjassot, València, Spain}

% ...



\begin{abstract}
Event Horizon Telescope (EHT) images of the supermassive black hole M87* depict an asymmetric ring of emission.  General relativistic magnetohydrodynamic (GRMHD) models of M87* and its accretion disk predict that the amplitude and location of the ring's peak brightness asymmetry should fluctuate due to turbulence in the source plasma.  We compare the observed distribution of brightness asymmetry amplitudes to the simulated distribution in GRMHD models, across varying black hole spin $\spin$.  We show that, for strongly magnetized (MAD) models, three epochs of EHT data marginally disfavor $|\spin| \lesssim 0.2$. This is consistent with the Blandford–Znajek model for M87's jet, which predicts that M87* should have nonzero spin.  We show quantitatively how future observations could improve spin constraints, 
and discuss how improved spin constraints could distinguish between differing jet-launching mechanisms and black hole growth scenarios.

\end{abstract}


\section{Introduction}

The Event Horizon Telescope (EHT) has captured images of the supermassive black hole M87* \citep{EHTC_2019_1, EHTC_2024_1, EHTC_2025_1_M87_21}, at the center of the M87 galaxy. These images feature a bright ring-like structure (the ``ring'') that encloses a dark low-intensity region (the ``shadow''). The ring is produced by synchrotron emission from hot plasma located within a few Schwarzschild radii of the black hole. The shadow corresponds to the region where the observer's line of sight intersects the event horizon.  The ring diameter is consistent with a black hole of mass $M = 6.5 \times 10^9 M_\odot $ located at a distance $D = 16.8$ Mpc from Earth \citep{EHTC_2019_6}, in agreement with stellar kinematics measurements \citep{Gebhardt_2011, Liepold_Chungpei_2023}. 
The ring has a north–south brightness asymmetry, with the brightness peak shifting to the southwest in recent observations \citep{EHTC_2024_1, EHTC_2025_1_M87_21}.

\begin{figure}
    \centering
    \includegraphics[width=0.8\textwidth]{ 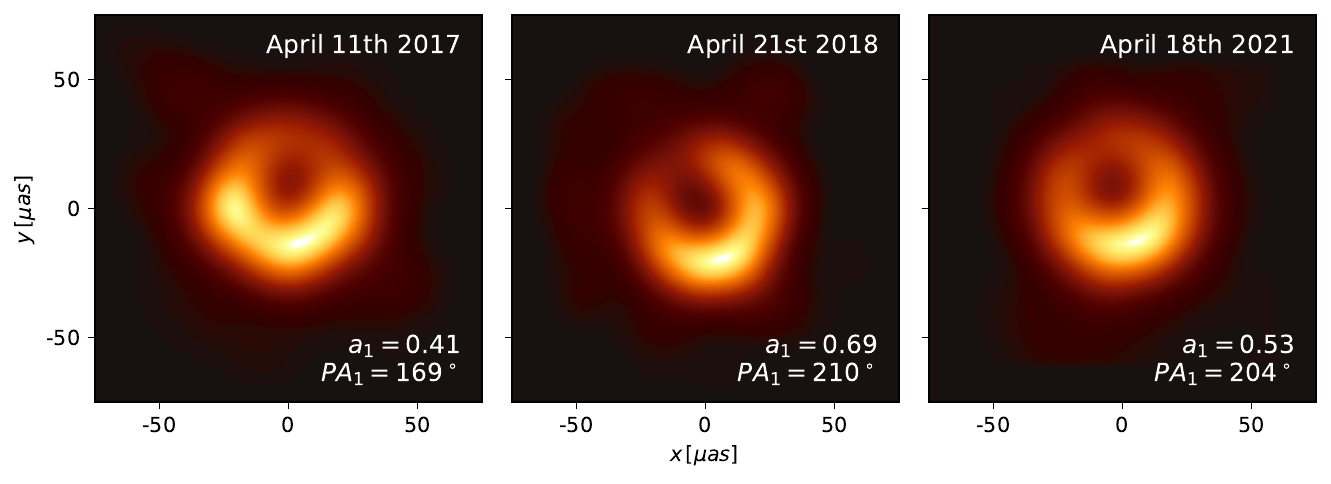}
    \caption{EHT images of M87* on April 11, 2017  \citepalias[][left]{EHTC_2019_1},  April 21, 2018
    \citepalias[][middle]{EHTC_2024_1}, and April 18, 2021 \citepalias[][right]{EHTC_2025_1_M87_21}. All images show a brightness asymmetry with a peak magnitude in the south.  
    The asymmetry magnitude $\asym$ and angle $\mathrm{PA}_1$ measured in this work are shown.
    }
    \label{fig:EHT_images}
\end{figure}

In this paper, we study how ring brightness asymmetry constrains dimensionless black hole spin: $\spin \equiv Jc/GM^2$ for black hole angular momentum $J$, speed of light $c$, and gravitational constant $G$. 
The task of recovering robust spin measurements for supermassive black holes is a long-standing problem in the field. 
From a theoretical perspective, spin and mass are the two quantities that completely define 
the spacetime 
geometry of any uncharged black hole \citep{kerr_1963}. In the astrophysical context, spin can be used to discriminate different scenarios of black hole formation and growth \citep[e.g.,][]{merloni_2008, bambi_2021, reynolds_2021, ricarte_2023}. 
Tilt between the spin axis and the accretion disk angular momentum axis can lead to Lense-Thirring precession of the disk and jet \citep[e.g.,][]{Bardeen_1975_tilt, Fragile_2005_tilt, Fragile_2007_tilt, Shiokawa_thesis_tilt, Liska_2018_tilt}.
Spin is also predicted to affect orbits of stars passing close to the black hole via Lense-Thirring precession \citep{levin_2003} and resonant friction \citep{levin_2024} effects. It can even be used to constrain merger history and cosmological formation of the host galaxy \citep[e.g.,][]{gammie_2004, ricarte_2023, ricarte_2025, sala_2024}.

Determining spins of supermassive black holes is especially important for understanding jet formation. To this day, the physical mechanism of jet launching has not been confirmed.
The Blandford–Znajek mechanism \citep{blandford_znajek}, which is the leading model for jet formation, predicts that jets are powered by frame dragging of magnetic field lines that thread the horizon of a spinning black hole. The extraction of black hole spin energy powers the jet. There are alternatives, however. One of the most prominent competing models is the Blandford-Payne mechanism \citep{blandford_payne}, in which accretion disk rotation powers the jet. Other models suggest gas, radiation \citep{rees_1981}, and magnetic field \citep{lynden_bell_1994} pressures as alternative driving mechanisms. 
These models predict different probability distributions for the spin based on the presence of a jet with a given power. They likewise predict different equilibrium values for spin over cosmic time. Therefore, precise measurement of spin is important  for determining the correct model of jet launching. Such an analysis is relevant for both M87*, where the jet feature is prominent \citep[see, e.g.,][]{walker_jet_2018}, and Sgr A*, where the presence of a jet is still debated.  

As of now, spin remains elusive. Current spin constraints rely on X-ray techniques that apply only to high accretion rate systems  \citep[$0.01 \lesssim \dot{M}/\dot{M}_{\text{Edd}} \lesssim 0.3$, per][]{reynolds_2021, bambi_2021}.  No observational spin constraints exist for lower accretion black holes ($\dot{M}/\dot{M}_{\text{Edd}} < 0.01$), which are far more prevalent in the local universe (\citealp{merloni_2008}; although many theoretical studies of spin evolution in these systems have been made, including \citealp{pacucci_2020, ricarte_2023, sala_2024}).
The spin of Sgr A* in our own galactic center, for example, is a topic of current research.

For M87*, there is currently no direct, statistically significant spin constraint. Indirect spin constraints have been made via the jet power, favoring $|\spin|>0$  \citep[e.g.][]{Nemmen_2019_jetpower, EHTC_2019_5}. This analysis, however, relies on highly uncertain observational constraints, assumes a Blandford–Znajek powered jet, and depends on jet launching at an earlier time. This motivates a conservative approach for GRMHD model-dependent spin constraints with jet power \citep{EHTC_2019_5}. 
Near-horizon electric vector polarization angles also marginally favor retrograde spins, where the accretion angular momentum vector and the black hole spin vector are antialigned \citep{palumbo_2020, EHTC_2021_8, EHTC_2025_1, janssen_2025}. However, uncertainties in interpreting polarization data (due to, e.g., Faraday rotation and year-to-year variability) motivate complementary constraints using near-horizon total intensity data.

Here, we constrain spin by comparing a particular measure of the brightness asymmetry magnitude (using the amplitude of the fitted $m=1$ $m$-ring) in EHT observations of M87* with the same measure in a library of synthetic data based on GRMHD simulations.  Our library---the ``Illinois v5'' \kharma library---contains models with a dense distribution of spins and parameterized electron distribution functions.  The observational data were obtained by the EHT Collaboration during the 2017, 2018, and 2021 observing campaigns (hereafter EHTC M87 I–VI for \citealt{EHTC_2019_1,EHTC_2019_2,EHTC_2019_3,EHTC_2019_4,EHTC_2019_5,EHTC_2019_6}; EHTC M87 2018 I–II for \citealt{EHTC_2024_1,EHTC_2025_1}; and EHTC M87 2021 for \citealt{EHTC_2025_1_M87_21}).
Reconstructed EHT images from all three years are shown in Figure \ref{fig:EHT_images}.

The observed ring in Figure \ref{fig:EHT_images} has a brightness asymmetry. Earlier studies of asymmetry in M87* have considered its dependence on the black hole inclination \citep[see][]{medeiros_asym_2022}, and compared it to the ring asymmetry of Sgr A* \citep[see][]{faggert_asym_2024}.  The brightness asymmetry is produced by a complicated interplay of Doppler boosting, gravitational lensing, frame dragging, radiative transfer effects, and gravitational redshift \citepalias{EHTC_2019_5}, which can depend on spin.  It is possible, therefore, that there exists a correlation between the asymmetry magnitude and the spin of the black hole.  This paper investigates that correlation using GRMHD models.  

The paper is structured as follows. In Section \ref{sec:models}, we describe a strongly magnetized (MAD) model library. Section \ref{sec:methodology} discusses how brightness asymmetry is estimated from the data and from synthetic images. In Section \ref{sec:results}, we present the distribution of asymmetry across spin. We then discuss in Section \ref{sec:discussion}, 
including a comparison to earlier work (\ref{subsec:comparison}), an analysis of future observations  (\ref{subsec:future_observations}), a discussion of limitations  (\ref{sec:incl} to \ref{sec:uncert}), and a proposed test of the Blandford–Znajek mechanism (\ref{subsec:bz}). Finally, we summarize in Section \ref{sec:conclusion}. In the Appendix, we investigate the distribution of asymmetries in weakly magnetized (SANE) models and prior GRMHD models  (Appendix \ref{sec:appendix}), demonstrate that our spin constraints are insensitive to methodology changes (Appendix \ref{subsec:appendix_bounds_coverage}), and present fits to the distribution of asymmetry in all models (Appendix \ref{subsec:appendix_fits}).

\section{Models}
\label{sec:models}

To assess the probability of different spins producing the observed asymmetry, we compare EHT 
data to a library of synthetic data based on GRMHD models which span a range of model parameters. 

The library was produced using the \patoka pipeline \citep{wong_patoka_2022}, which models the accretion flow using ideal general relativistic magnetohydrodynamics (GRMHD) simulations and then ray-traces GRMHD snapshots to produce synthetic images. The GRMHD simulations are integrated with the \kharma code \citep{Prather_kharma_2024}, and the synthetic images are produced using \ipole \citep{ipole18}.  For this study, we use the ``v5'' version of the Illinois GRMHD library (Bowden et al., in preparation).  Results from the v3 library \citep{Dhruv_grmhd_survey_2024}, which were produced with an outdated adiabatic index among other differences, used the \iharm code \citep{prather_iharm_2021} and are described in Appendix \ref{sec:appendix}.  

Our GRMHD simulations are a natural progression of the radiatively inefficient accretion flow (RIAF) model \citep{Ichimaru_1977_riaf, Narayan_1994_riaf, Narayan_1995_thick2, Quataert_1999_riaf, yuan_2003_riaf}. These accretion disks are typically optically thin and geometrically thick, comprised of relativistically hot plasma \citep{Rees_1982_thick, Naryana_1995_thick, Narayan_1995_thick2, Reynolds_1996_thick, Yuan_2002_thick, DiMatteo_2003_thick, Yuan_2014_thick}. This model is well-supported by observations, which find that M87 has a relatively low-luminosity active galactic nuclei, suggesting that M87* has a low accretion rate \citep[e.g.,][]{Yuan_2014_thick}. The RIAF thick-disk model is also supported by EHT observations \citepalias{EHTC_2019_5}. See \citet{wong_patoka_2022} for additional discussion.

Our GRMHD fluid simulations have two parameters: accretion state and black hole spin.  The accretion state typically falls into strongly or weakly magnetized modes (MAD or SANE). Magnetically arrested disk (MAD) models \citep{igumenshchev03, narayan03, Tchekhovskoy_2011} have near-horizon magnetic fields that are strong enough to impede accretion.  This limits the magnetic flux on the horizon to a characteristic value that depends on the accretion rate. Standard and normal evolution (SANE) models \citep{narayan12, sadowski_2013} have weaker near-horizon fields and are more similar to classical thin disks, although they are optically thin and geometrically thick. In the main text, we focus on MAD models.  This is motivated by evidence that MAD models are favored by M87* data, as they are more likely to pass jet power and polarization constraints \citepalias{EHTC_2019_5, EHTC_2023_9}. For v5, we densely sample the dimensionless spin $\spin$ in 13 steps, ranging from $-0.97$ to $0.97$ (Bowden et al., in preparation).
Recall that for a Kerr black hole, $\spin\in[-1,1]$. Here, positive spin means the accretion disk angular momentum vector is prograde (aligned) with respect to black hole spin vector, while negative spin means that the accretion disk angular momentum vector is retrograde (antialigned). The spin and accretion disk angular momentum vector are assumed to be coaxial: we do not consider tilted disks. Our analysis therefore uses 13 MAD GRMHD fluid simulations.  

The radiative transfer step has three parameters: inclination, $\rhigh$, and $\rlow$.

The inclination $i$ is the angle between the accretion flow orbital angular momentum vector and the line of sight.  We assume that the black hole spin vector, the accretion flow orbital angular momentum vector, and the large scale jet are coaxial, and that the approaching jet inclination is $17^\circ$ \citep[see, e.g.,][]{walker_jet_2018}.  The position angle of the brightest point on the ring is known to depend on spin, and the position angle of the observed asymmetry implies that the spin vector is pointed away from Earth \citepalias{EHTC_2019_5}.  Adopting this constraint, we set $i = 17^\circ$ for $\spin < 0$ and $i = 163^\circ$ for $\spin \ge 0$.  The different inclination angle for positive and negative spins has no impact on the asymmetry magnitude distribution.

The radiative properties of the plasma depend on the electron distribution function, which is not determined by the GRMHD simulations.  We use a model with parameters $\rhigh$ and $\rlow$ to assign a distribution function to the electrons.  The model assumes that (1) the electrons are thermal at temperature $T_e$; (2) the ions are thermal at temperature $T_i$; (3) $T_i$ can be deduced directly from the simulation; and (4) $T_i/T_e$ depends only on the local, instantaneous value of the plasma $\beta \equiv P_{gas}/P_{mag}$ for gas pressure $P_{gas}$ and magnetic pressure $P_{mag}$.  Then, following  \cite{moscibrodzca16}, 
\begin{equation}\label{eq_edf}
    \frac{T_i}{T_e} = \rhigh \frac{\tilde{\beta}^2}{1+\tilde{\beta}^2} + \rlow \frac{1}{1+\tilde{\beta}^2}
\end{equation}
for $\tilde{\beta}=\beta/\beta_{\text{crit}}$ and $\beta_{\text{crit}}=1$. $\rhigh$ and $\rlow$ are thus dimensionless temperature ratios.
$\rlow$ dominates at low $\tilde{\beta}$ and $\rhigh$ at high $\tilde{\beta}$.
We consider models with $R_{\text{high}} = 10, 40, 80, 160$ and $\rlow = 1, 10$ \citepalias{EHTC_2019_5, EHTC_2021_8}. 
We do not consider $\rhigh=1$ because it is disfavored by the data \citepalias{EHTC_2019_5, EHTC_2023_9, EHTC_2025_1}.
There are thus 8 unique combinations of $i$, $\rhigh$, and $\rlow$ .

\begin{figure}[t]
    \centering
    \includegraphics[width=1.0\textwidth]{ 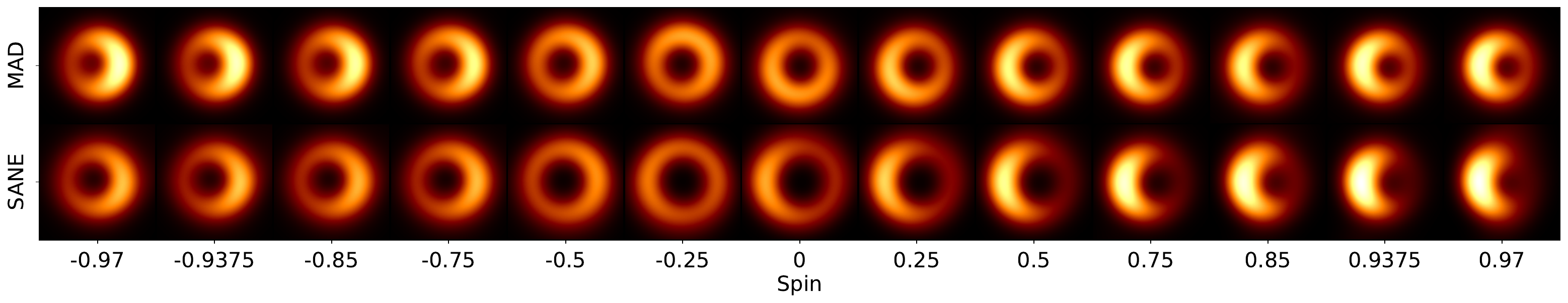}
    \caption{
    Mean images for all models in the v5 library (averaging over  $\rhigh$, $\rlow$, and time, and blurred using a $15 \, \mu as$ full width at half maximum Gaussian kernel) at $i=17^\circ$ for $a_{*} < 0$ and $i=163^\circ$ for $a_{*} \geqslant 0$, across each spin (columns) for MADs (top row) and SANEs (bottom row).  Asymmetry magnitude typically increases with $|\spin|$.  In these images, the projected prograde spins point up, retrograde spins point down, the accretion flow angular momentum points up, and jet axis is vertical. The position angle of the brightness asymmetry approximately follows the sign of the spin, consistent with \citepalias{EHTC_2019_5}. See Appendix \ref{sec:appendix} for a discussion of SANEs.
    } 
    \label{fig:mean_img}
\end{figure}

Model parameters are listed in Table \ref{tab:model_params}. Each of 104 \patoka models contains 6000 images, produced at $5 \, t_g$ cadence (for gravitational time $t_g\equiv GMc^{-3}$) in the interval from time $20,000 \, t_g$ to $50,000 \, t_g$ after the start of the simulation. Time-averaged images for different spins are shown in Figure \ref{fig:mean_img} for both MAD and SANE models. It is clear from the figure that ring asymmetry and position angle are correlated with spin: as the magnitude of spin increases, the magnitude of the brightness asymmetry generally also increases. 

A comparison of MAD and SANE models in our main v5 library and in the v3 library, which is run with different physical assumptions and a different GRMHD code, are discussed in Appendix \ref{sec:appendix}. We prefer v5 models because (1) they are run at higher resolution; (2) they are run longer and therefore provide better sampling of the asymmetry distribution; (3) they sample spin more densely (13 spins in v5 vs 5 spins in v3); and (4) v5 uses adiabatic index $\gamma = 5/3$, while v3 uses $\gamma = 4/3$.  If $T_e \ll T_i$, then the gas pressure is dominated by the ions, which are nonrelativistic, and thus $\gamma = 5/3$ is more appropriate \citep[for a discussion, see][]{Gammie_2025}.
In the following section, we discuss how we measure asymmetry in models and observational data.

\begin{deluxetable}{cc}
    \tablecaption{Model parameters 
    \label{tab:model_params}}
    \tablehead{\colhead{Parameter} & \colhead{Sampled Values} }
    \startdata
    Magnetic Flux Mode & MAD \\
        Spin & $0$, $\pm 0.25$, $\pm 0.5$, $\pm 0.75$, $\pm 0.85$, $\pm 0.9375$, $\pm 0.97$  \\
        $\rhigh$ & $10$, $40$, $80$, $160$ \\
        $\rlow$ & $1$, $10$ \\
        $i$ [deg] & $17^{\circ}$ (for $\spin < 0$), $163^{\circ}$ (for $\spin \geqslant 0$) 
    \enddata
    \tablecomments{v5 model parameters used for the analysis. Each combination of parameter values corresponds to one model. SANE models are discussed in Appendix \ref{sec:appendix}. }
\end{deluxetable}

\section{Measuring Asymmetry}
\label{sec:methodology}

\subsection{EHT observations} \label{method:observations}

Very-long-baseline interferometry (VLBI) arrays such as the EHT sample the $(u,v)$, or Fourier, domain. The sampling is based on the coverage of their baselines projected onto the sky.
To measure ring asymmetry with $(u,v)$ data, we fit a parameterized $m$-ring model using the Comrade package for Bayesian modeling of VLBI observations \citep{tiede_comrade_2022}.
We demonstrate consistency with asymmetry fits in the image domain using variational image domain analysis \citep[VIDA;][]{tiede_vida_2022} in Section \ref{sec:domain}.  Detailed validation of the parameter optimization method is described in \cite{tiede_vida_2022}.  

Our parameterized Comrade model has two components: a ring with intensity $I_{ring}$ and a Gaussian background component with intensity $I_{bg}$.  The total intensity is $I = I_{ring} + I_{bg}$.  The Gaussian background intensity is
\begin{equation}\label{eq_background}
I_{bg}(x, y) \propto (1- f) \exp\left(-\frac{(x')^2}{2\sigma_g^2}-\frac{(y')^2}{2\sigma_g^2(1+\tau_g)^2}\right)
\end{equation}
for $x'=(x-x_g)\cos \xi_g +(y-y_g) \sin\xi_g$ and $y'=-(x-x_g)\cos \xi_g +(y-y_g) \sin\xi_g$. 
The Gaussian is scaled by $(1-f)$ with standard deviation $\sigma_g$, given ellipticity $\tau_g$, rotated by angle $\xi_g$, and shifted by $(x_g,y_g)$. The background component thus has 6 parameters  (these parameters are unitless, as the fitting is done to normalized images; the images can be rescaled after fitting as needed). This Gaussian background component is typical for modeling EHT data (\citealp{tiede_vida_2022}; \citetalias{EHTC_2025_1}). The ring intensity is modeled with a so-called $m=4$ $m$-ring:
\begin{equation}\label{eq_ring_model}
I_{ring}(r, \theta) \propto \left[ 1 - \sum_{m=1}^{4} a_m \cos \left( m \theta - \mathrm{PA}_m \right) \right] \exp \left( -\frac{(r-r_0)^2}{2\sigma^2} \right).
\end{equation}
The $m$-ring is a Fourier series wrapped along a ring of Gaussian thickness \citep{johnson_photon_2020, tiede_vida_2022}. We truncate the Fourier series at $m = 4$ \citepalias[which is typically sufficient to describe EHT sources;][]{EHTC_2024_1} and require that $a_m < 1$ to ensure positive intensity everywhere. Our $m=4$ $m$-ring has 10 parameters: ring radius $r_0$, ring thickness $\sigma$, amplitudes $a_m$, and position angles $\textrm{PA}_m$ for $m=1$ through $4$. Here, $\asym$ captures the large-scale brightness asymmetry, while $a_2$ through $a_4$ capture smaller scale turbulent structure.
We use $\asym$, the normalized amplitude of the $m = 1$ Fourier component, as our measure of ring brightness asymmetry. In our convention, notice that a ring with $\asym=0$ is symmetric, while a ring with $\asym=1$ has an amplitude as large as its mean.

The model parameters are adjusted by the Comrade optimizers to fit EHT visibility amplitude and closure phase data. Using this procedure to model EHT 2017, 2018, and 2021 observations \citepalias{EHTC_2019_1, EHTC_2024_1, EHTC_2025_1_M87_21} yields $\asym = 0.41 \pm 0.04$, $\asym = 0.69 \pm 0.10$, and
$\asym = 0.53 \pm 0.04$ respectively. We use a different measure of asymmetry than 
\citetalias{EHTC_2019_4}, \citetalias{EHTC_2024_1}, and \citetalias{EHTC_2025_1_M87_21}, and therefore the  asymmetries reported here differ from prior values reported by EHT (See Section \ref{subsec:comparison}). The quality of fit can be assessed using the  mean squared standardized residual $\chi^2$, normalized by the number of data points. The normalized $\chi^2$ is $1.9$, $0.9$, and $2.6$ in 2017, 2018, and 2021 respectively (note these are not reduced $\chi^2$ values). For additional discussion, see \citetalias{EHTC_2019_4}, \citet{tiede_comrade_2022}, and \citetalias{EHTC_2024_1}, which finds that an $m=4$ $m$-ring has among the highest Bayesian evidence when fitting to EHT data in a survey of 25 geometric modeling templates. Our results are insensitive to changes in the $m$-ring modeling template, as described in Appendix \ref{subsec:appendix_bounds_coverage}.

\subsection{Simulated images}
\label{sec:domain}

We apply the same analysis technique to our synthetic data as we did to the real data.  Each GRMHD model contains a sequence of images separated by $5 \, t_g$.  The images are correlated on timescales $\sim 100 \, t_g$, so we downsample to a cadence of $125 \, t_g$ (see Section \ref{subsec:future_observations}) to obtain independent synthetic observations. This leaves $240$ images per unique combination of spin, $\rhigh$, and $\rlow$, and $1920$ unique images per unique spin. 

To measure asymmetry in models, we require synthetic data generated with realistic observational conditions. We first rotate the ground truth images to match the large-scale jet axis expected from observations (again, we assume the large-scale jet, disk angular momentum vector, and black hole spin vector are coaxial). We generate $(u,v)$ data by Fourier transforming our synthetic images, then sampling the resulting visibility amplitudes and closure phases using EHT $(u,v)$ coverage, with realistic observational noise \citep[including baseline-dependent thermal noise, generated using the eht-imaging software;][]{Chael_2018}. Finally, we fit the $m$-ring model parameters to the resulting synthetic data using Comrade, as in Section \ref{method:observations}. 
For simplicity, we assume 2018 EHT $(u,v)$ coverage and conditions.  The results are consistent if we use 2017 or 2021 $(u,v)$ coverage and observational noise instead (see Appendix \ref{subsec:appendix_bounds_coverage}), or fit directly in the image domain.

\begin{figure}[t]
    \centering
    \includegraphics[width=0.8\textwidth]{ 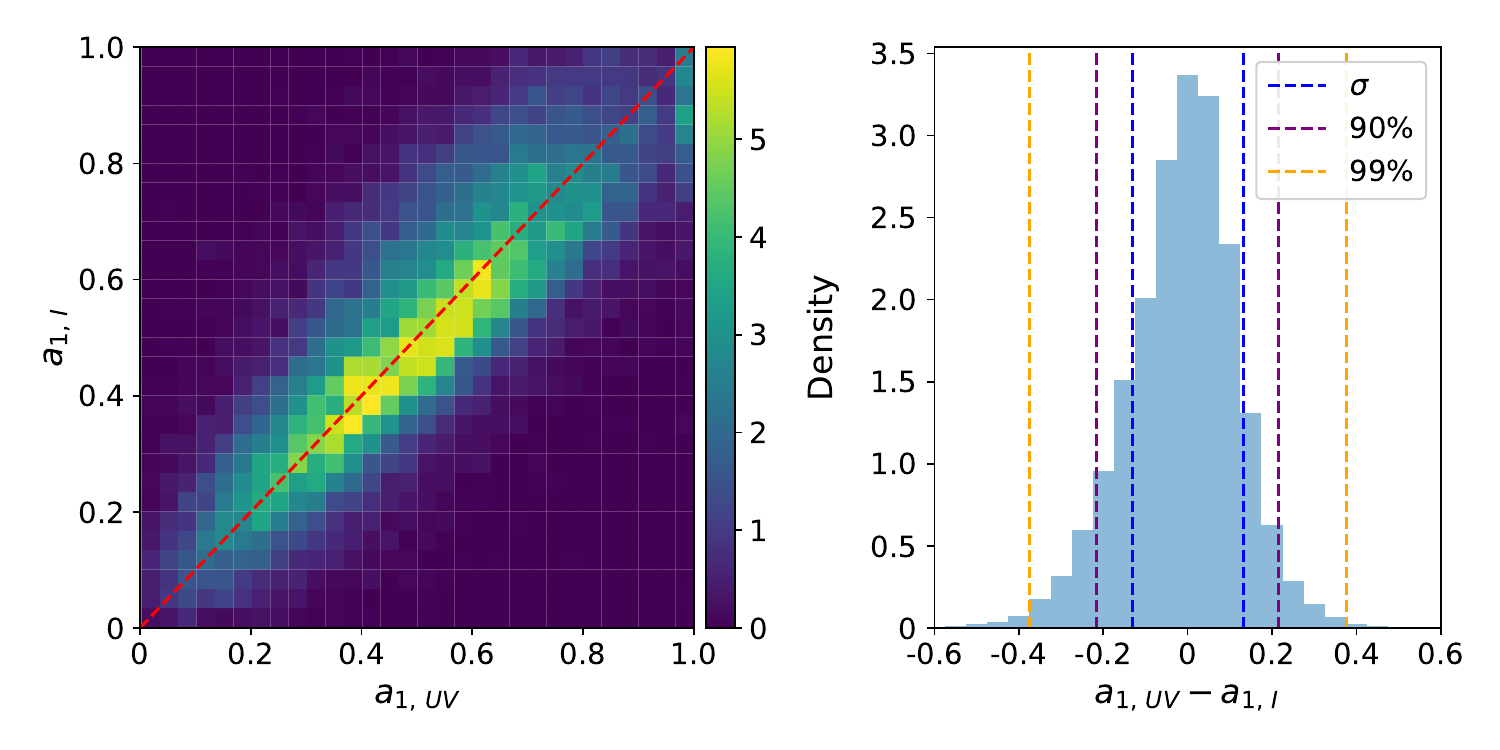}
    
    \caption{Comparison of the asymmetry magnitude measured in the image domain  $a_{1,I}$ and $(u,v)$ domain $a_{1, UV}$ for MAD models, plotted as the density of total snapshots. Left: the distribution of snapshots in $(a_{1,I}, \,a_{1, UV})$ space. The red line corresponding to $a_{1,I} = a_{1, UV}$. We set a ceiling of $a_{1, UV} \leq 1$ in the $(u,v)$ optimizer, which accounts for the cluster of snapshots at $a_{1, UV}=1$. These outlier values occur in $<1\%$ of total snapshots and do not affect our results (see Appendix \ref{subsec:appendix_bounds_coverage}). Right: distribution of $a_{1, UV} - a_{1,I}$, with $1\sigma$, $90\%$, and $99\%$ boundaries marked. Nearly $90\%$ of snapshots have $|a_{1, UV} - a_{1,I}| < 0.2$.}
    \label{fig:uv_img}
\end{figure}

\begin{figure}[t]
    \centering
    \includegraphics[width=0.8\textwidth]{ 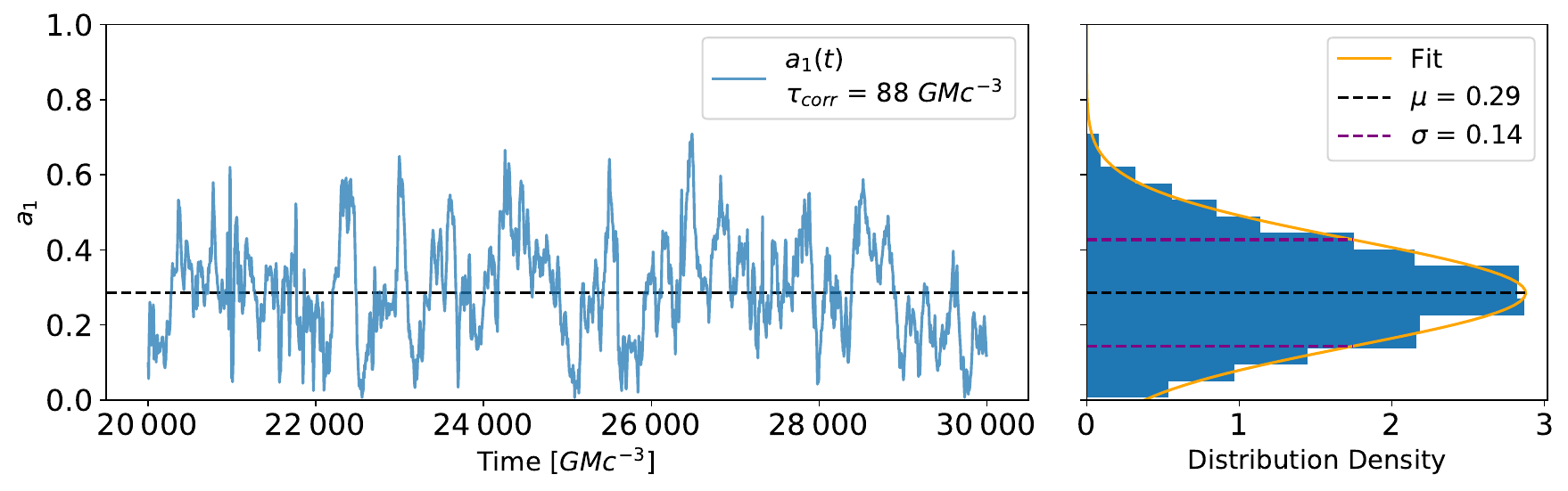}
    \caption{Left: evolution of asymmetry $\asym$ in time, over a 10,000 $t_g$ time window, for the $\spin = 0$, $\rlow=1$, $\rhigh=40$, MAD model. Right: the distribution of asymmetry amplitudes for this model window. The truncated Gaussian fit to the distribution is shown, with the mode $\mu = 0.29$, and standard deviation $\sigma = 0.14$. }
    \label{fig:asym_evol}
    
\end{figure}

A comparison of asymmetry $a_{1,I}$ measured in the image domain to asymmetry $a_{1, UV}$ measured from synthetic $(u,v)$ data shows the effect of limited $(u,v)$ coverage.  We have compared $a_{1, UV}$ and $a_{1,I}$ for all snapshots in the library.  The results are shown in Figure \ref{fig:uv_img}.  About $68\%$ of snapshots have $|a_{1, UV} - a_{1,I}|$ less than $0.13$, and $90\%$ have difference less than $ 0.22$.  Furthermore, the root mean squared difference is smaller than the variation of asymmetry in an individual model.  Evidently $a_{1, UV}$ and $a_{1,I}$ are highly correlated and $a_{1, UV}$ provides a nearly unbiased estimate of $a_{1,I}$ (notice the minimal offset in Figure \ref{fig:uv_img}). 
A small fraction of snapshots cluster around $a_{1, UV}=1$. This artifact occurs due to the chosen bounds and does not effect the results.
We conclude that asymmetry can be accurately measured, even with EHT's limited $(u,v)$ coverage. 

The asymmetry fluctuates over time. Across the v5 model library, we find a mean correlation timescale for $\asym$ of $\tau_{corr}\approx 70\,t_g$ for SANEs and $\tau_{corr}\approx 115\,t_g$ for MADs. An example asymmetry evolution for a model with $\spin = 0$, $\rlow = 1$, $\rhigh = 40$ is shown in Figure \ref{fig:asym_evol}.  The asymmetry approximately follows a truncated normal distribution.  For this example, the $\asym$ distribution has mode $\mu = 0.29$, standard deviation $\sigma \approx 0.14$, and correlation timescale $ 88 \, t_g$. 

\section{Results}
\label{sec:results}

\begin{figure}[t]
    \centering
    \includegraphics[width=0.5\textwidth]{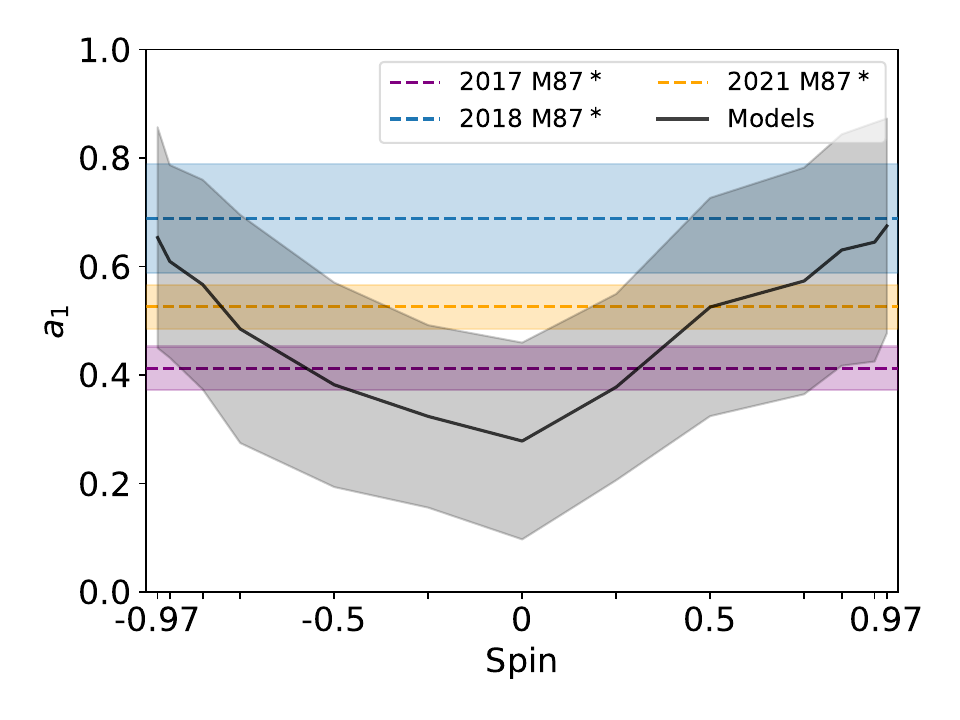}
    \caption{Distribution of MAD model asymmetry. The black line shows the modes $\mu$ of truncated Gaussians fitted to modeled asymmetry distributions, marginalizing over $\rhigh$ and $\rlow$, at each spin. The gray band shows $1\sigma$ distribution variation. EHT  measurements of asymmetry in M87* are shown as dashed lines, surrounded by $1\sigma$ regions (colored shaded bands). Spin values present in the models are shown with x-axis tick marks.
    The 2017, 2018, and 2021 data fall above the 50th percentile for zero-spin models. The 2018 asymmetry falls above the 90th percentile for zero-spin models.}
    \label{fig:spin_distr}
\end{figure}

After measuring asymmetry for all MAD models, we find the asymmetry distributions show a clear spin dependence, consistent with Figure \ref{fig:mean_img}.  Figure \ref{fig:spin_distr} shows the asymmetry distributions for MAD models, marginalizing over all parameters except spin. 
Models with zero spin show minimum asymmetry, while models with maximum spin magnitude have mean asymmetry that is over twice as large.  
Truncated Gaussian fit parameters to these model distributions are provided in Table \ref{tab:asym_gauss_fits}. Fits for SANE models are provided in the Appendix \ref{subsec:appendix_fits}.
We filter out all images with $\asym > 0.99$ before fitting to improve accuracy. We note that observations are well below this threshold, and that this process has no major effect on our statistical results (see Appendix \ref{subsec:appendix_bounds_coverage}).

The figure also shows the observed asymmetries from 2017 to 2021.
As above and in Figure \ref{fig:EHT_images}, the measured asymmetry for 2017, 2018, and 2021 EHT data are $0.41$, $0.69$, and $0.53$ respectively. The asymmetry is largest in 2018, when there appears to be the largest difference between the ring's brightness peak and brightness minimum. The 2021 image has a similar position angle, but has relatively more emission on the ring's dim side. 2017 has the least asymmetry, with a smaller difference between the ring's maximum and minimum brightness, and with a wider multimodal bright side. Our measured asymmetry values fit what is seen by-eye. 
In Figure \ref{fig:spin_distr}, we see the 2018 asymmetry is above the 90th percentile of all non-spinning black hole models, while the 2017 and 2021 asymmetries lie above the 50th percentile of all non-spinning black hole models.  The data clearly disfavor zero-spin models.

\begin{deluxetable}{cccccccccccccc}
    \label{tab:asym_gauss_fits}
    \tablecaption{Gaussian Fits to Asymmetry}
    \tablewidth{\textwidth}
    \tablehead{
    \colhead{Spin} & \colhead{$-0.97$} & \colhead{$-0.9375$} & \colhead{$-0.85$} & \colhead{$-0.75$} & \colhead{$-0.5$} & \colhead{$-0.25$} & \colhead{$0.0$} & \colhead{$0.25$} & \colhead{$0.5$} & \colhead{$0.75$} & \colhead{$0.85$} & \colhead{$0.9375$} & \colhead{$0.97$} }
    \startdata
    $\mu$
    & $0.653$ & $0.609$ & $0.567$ & $0.485$ & $0.382$ & $0.324$ & $0.278$ & $0.378$ & $0.525$ & $0.573$ & $0.630$ & $0.645$ & $0.675$ \\
    $\sigma$
    & $0.203$ & $0.178$ & $0.193$ & $0.210$ & $0.188$ & $0.168$ & $0.181$ & $0.172$ & $0.201$ & $0.209$ & $0.213$ & $0.220$ & $0.198$
    \enddata
    \tablecomments{Mode ($\mu$) and standard deviation ($\sigma$) for truncated Gaussian fits to asymmetry distributions for MAD models. 
    The fits are bounded by $\asym \in [0, 1)$; we exclude $\asym = 1$ cases prior to fitting.}
\end{deluxetable}

To assess the significance of the inconsistency between the data and the models, we use the Kolmogorov–Smirnov (KS) test.  The KS test provides a probability $p$ that the difference between the observed and model cumulative distributions are as large or larger than observed.  

\begin{figure}[t]
    \centering
    \includegraphics[width=0.5\textwidth]{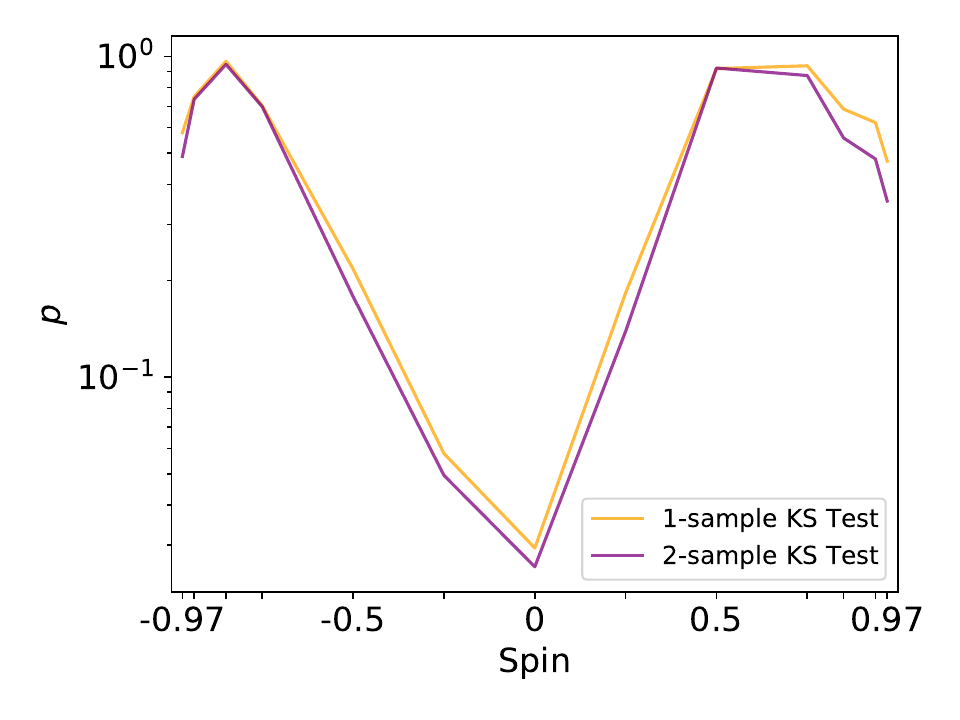}
    \caption{Probability $p$ that the ensemble of MAD models at fixed spin is consistent with EHT data. The purple curve shows the results from the two-sample KS test against the full asymmetry distributions obtained for each spin. The yellow curve shows one-sample KS tests, with underlying asymmetry distributions replaced with best-fit truncated normal distributions. Parameters of these distributions are given in Table \ref{tab:asym_gauss_fits}. The results are largely consistent.}
    \label{fig:p_values}
\end{figure}

The probability for each spin, marginalizing over each model $\rhigh$ and $\rlow$, are shown in Figure \ref{fig:p_values}.  There are two lines in the figure: one for the truncated Gaussian fit to the GRMHD asymmetry distribution (the ``one-sample'' KS test), and one using measured asymmetries of individual images drawn from each model (the ``two-sample'' KS test).  Evidently the two are very close.   Models with $\spin = 0$ and $\spin = -0.25$ give lower p-values than higher spin models: $\spin=0$ models have $p=0.026$ and $\spin=-0.25$ models have $p=0.049$ in the two-sample KS test. Zero spin models are strongly disfavored. Maximum $p$-values occur at $\spin=-0.75, 0.5,$ and $0.75$, as spins with larger magnitudes produce asymmetries that slightly exceed the current mean of observations. However, the relative difference in $p$ between these highly spinning models is insignificant; all are consistent with the data at present. 
If the spin of M87* is nonzero and the models provide the correct asymmetry distributions, then future observations will soon tighten these constraints.  

\begin{figure}[h]
    \centering
    \includegraphics[width=0.5\textwidth]{ 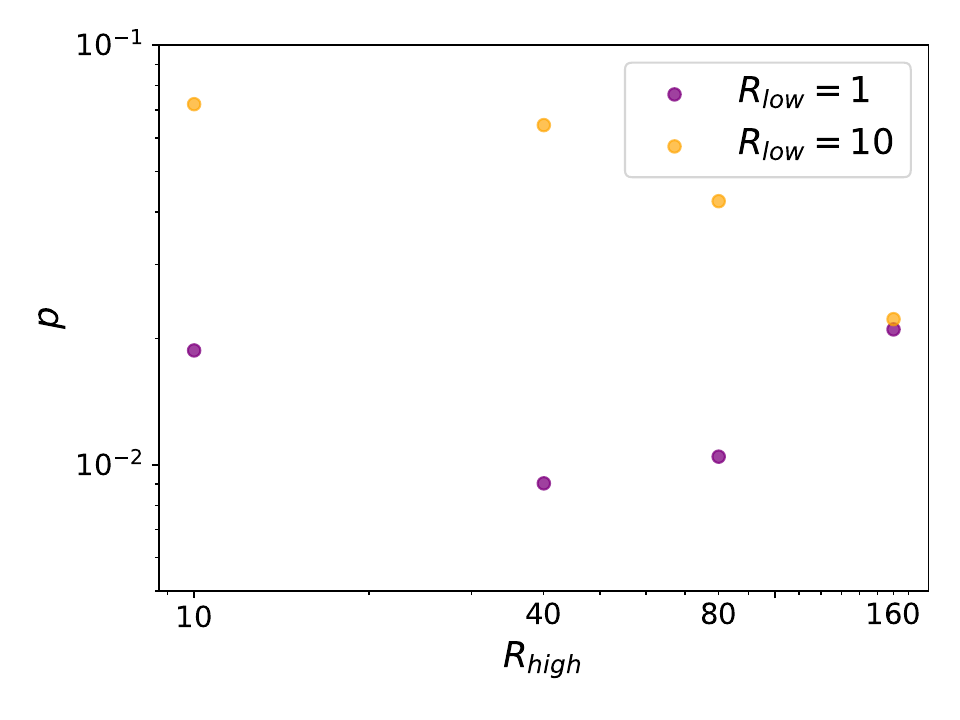}
    \caption{Probability $p$ for zero-spin models, across $\rlow = 1$ (purple) and $\rlow = 10$ (yellow), based on two-sample KS tests. Models with $\rlow = 10$ are marginally favored.}
    \label{fig:p_value_spin_0}
\end{figure}

In the discussion above, we marginalize over all radiative transfer parameters $\rhigh$ and $\rlow$ at each spin.  Figure \ref{fig:p_value_spin_0} shows KS test probabilities for $\spin = 0$ models broken out for different $\rhigh$ and $\rlow$.  Recall that these parameters set the ratio of electron temperature to ion temperature as a function of plasma $\beta$.  

The asymmetry is only weakly dependent on $\rhigh$ and $\rlow$ in  MAD models. The emission region is relatively insensitive to $\rhigh$ and $\rlow$ in  MADs (\citetalias{EHTC_2019_5}; \citealp{wong_patoka_2022}), so the asymmetry's weak dependence on these parameters is expected.
Still, the $\rlow = 1$ models tend to exhibit lower $p$ than $\rlow = 10$ models.  In Figure \ref{fig:p_value_spin_0}, we see this trend for $\spin=0$, as $\rlow=10$ models have slightly higher asymmetry than $\rlow = 1$, closer to observations. Prograde MAD $\rlow=10$ models produce slightly less asymmetry than $\rlow = 1$, also closer to the observed mean.
The $\rlow=10$ models are also favored by polarized EHT data: they more frequently reproduce the observed polarization fraction and structure. Cooler electron temperatures are a plausible consequence of radiative cooling.
Meanwhile, models with low $\rhigh$ (e.g. $\rhigh=1$) are disfavored by other constraints: they overproduce X-rays, underproduce jet power, are too radiatively efficient \citepalias{EHTC_2019_5}, and have a too high polarization fraction \citepalias{EHTC_2021_8, EHTC_2023_9}. While our asymmetry does not exclude $\rhigh=10$, it does mildly favor $\rlow=10$. It seems cooler electrons are preferred.

\section{Discussion}
\label{sec:discussion}

The asymmetry in black hole images is determined by multiple physical processes (Chael et al. 2025, in preparation).  These include Doppler beaming, gravitational lensing, and radiative transfer effects, e.g. the angle at which the line of sight passes through the midplane (lines of sight that are more nearly parallel to the midplane have higher opacity), and the angle between the magnetic field and the line of sight (lines of sight normal to the magnetic field have higher synchrotron opacity).   

Doppler shift is an important contributor to the asymmetry. Doppler shift depends on the emitting plasma velocity.  In Minkowski space, a rotating, optically thin ring will always appear brighter on the approaching side due to Doppler boosting: since the intensity $I_\nu$ on a line of sight and the frequency $\nu$ are invariant in the combination $I_\nu/\nu^3$, and $\nu$ is higher on the blueshifted side, then $I_\nu$ must be proportionately higher on the blueshifted side.  This argument incorrectly suggests, however, that the brightness peak should always lie on the approaching side of the disk. 

In the Kerr metric, a majority of the emission seen by EHT arises in a narrow ring close to the black hole at $r \sim 2-5 \, GMc^{-2}$ (\citetalias{EHTC_2019_5}; \citealp{wong_patoka_2022}). There, Doppler shift interacts with other effects such as lensing. For a rapidly spinning hole with equatorial emission seen at small inclination, lines of sight (geodesics) that intersect the emission region tend to bend and wrap around the hole in the spinward direction \citep{johnson_photon_2020}.  Thus for retrograde ($\spin < 0$) disks, direct lines of sight to the emission region emerge opposite the direction of rotation and are Doppler de-boosted.  Lensed, indirect lines of sight to the disk contribute comparatively more of the flux than in prograde disks.  

The strength of Doppler boosting depends on plasma angular velocity.  The angular velocity in the emission region is different in MAD (strongly magnetized) and SANE (weakly magnetized) models.  This is evident in Figure 9 of \cite{Dhruv_grmhd_survey_2024}, which shows time- and azimuth-averaged angular momentum in the emission region for both MAD and SANE models.  In MAD models, the plasma experiences strong magnetic torques in the plunging region, leading to slower rotation and thus weaker Doppler boosting.  Since MAD models have relatively weak Doppler boosting, the asymmetry can be more sensitive to other effects.

The rotational velocity also differs between v3 and v5 models.  This point is discussed in greater detail in Appendix \ref{sec:appendix}.  

\subsection{Comparison to Earlier Work}
\label{subsec:comparison}

EHT reported asymmetries for both the 2017 and 2018 M87* datasets \citepalias{EHTC_2019_4, EHTC_2024_1} that differ from those reported here.  Those reports were based on either different parameterized image models, different definitions of asymmetry, or both.  

The analysis of 2017 data reported an asymmetry $A\sim 0.22$ on April 11, which is smaller than reported here ($\asym=0.41$) due to a difference in definition. In that analysis, $A \equiv \langle|\int_{0}^{2 \pi} I e^{i\theta}d\theta|\,/\int_{0}^{2 \pi} I d\theta \rangle_{r\in[r_{in}, r_{out}]}$ on an annulus bounded by $r_{in}$ and $r_{out}$.  If we apply this definition for $A$ to an $m$-ring, the asymmetry recovered is a factor of 2 smaller than $\asym$, which is the  amplitude of the $m=1$ sinusoidal mode \citepalias[see Equation 22 and subsequent discussion in][]{EHTC_2019_4}.  The difference in definition accounts for most of the difference between their measurement and ours. The remainder is accounted for by a difference in measurement procedure: we fit a parameterized model, while the original EHT analysis did not. 

The EHT analysis of 2018 April 21 data used Comrade and VIDA to measure an asymmetry and found $A\approx 0.31$ \citepalias[other methods had the asymmetry ranging from $A\approx 0.21$ to $0.43$, with an average of $A\approx0.31$, per Table 7 of][]{EHTC_2024_1}. However, that analysis followed a Comrade convention where $m$-rings fitted in the $(u,v)$ domain produce amplitudes that are a factor of $2\times$ smaller than those fitted in the image domain. Thus, their values typically range from $a_{1, \, UV} \in[0, 0.5]$. In our convention, we present both $(u,v)$ and image domain asymmetries in units of $[\overline{I}_{ring}]$ ranging from $\asym\in[0,1]$, where $\asym = 1$ corresponds to a ring whose amplitude is as large its mean intensity $\overline{I}_{ring}$.

A smaller difference in asymmetry magnitude originates from the use of an  $mF$-ring or floored $m$-ring model in the EHT 2018 analysis. This model has a flat (non-Gaussian) background, an elliptical ring, up to $m=2$ modes (rather than $m=4$), and allows for two free nuisance Gaussians to help model diffuse extended emission. The nuisance Gaussians are allowed to overlap with the ring, and so can introduce on-ring brightness asymmetries that aren't accounted for in their measure of asymmetry $A$. The more complex template from 2018 analysis allows for better fits to the 2018 data, while ours allows for more straightforward interpretation at reduced computational cost.

The EHT analysis of 2021 April 18 used Comrade to measure an asymmetry of $A\approx 0.22$ to $0.25$, again following the convention where values of $A$ are a factor of 2 smaller than $\asym$. Their analysis used an $m=4$ $m$-ring as in this work, but had minor differences in the background Gaussian component \citepalias[see Section 3.4 and Table 2 in][]{EHTC_2025_1_M87_21}. Their recovered values of $A$ are consistent with $\asym$ measured in this work, after accounting for these differences.

The theoretical study of ring asymmetry in EHT images has focused mainly on inclination, rather than spin. \cite{medeiros_asym_2022} analyzed asymmetry in M87* with a focus on the inclination dependence, using a set of GRMHD models.  Their measure of asymmetry is based on integrations along rays through the image at fixed position angle, comparing the line-integrated intensity on one side of the shadow against the other.  Assuming that $a_3 \ll \asym$ (as in most of our fits), their asymmetry measure $A \approx (1 + \asym)/(1 - \asym)$.  They consider GRMHD models with $\spin = 0, 0.7, 0.9$ and see relatively little asymmetry at zero spin. \cite{medeiros_asym_2022} find a strong dependence on inclination, and also note a substantial difference between MAD and SANE models.

\cite{Qiu_2023} have used machine learning models trained on a different set of GRMHD simulations to assess relative importance of different image observables for spin and inclination inference. Asymmetry, computed as in \cite{medeiros_asym_2022}, was included as one of the observables. They find that asymmetry has moderate to strong effects on spin and inclination constraints, broadly consistent with our results.
The authors find that observation of asymmetry puts particularly strong constraints on spin for Sgr A* models. For M87* models, the constraining power of asymmetry is smaller, but is still the highest among all observables not pertaining to polarization structure. In our study, the v5 simulations sample black hole spin more densely and use what we believe to be a more accurate adiabatic index \citep{Gammie_2025}.

\cite{faggert_asym_2024} studied asymmetry in a phenomenological thick-disk model.  They use yet another measure of asymmetry, $FA$, which relies on the depth of nulls in the visibility amplitudes on rays perpendicular to the projection of the spin axis on the sky.  In the limit of a thin ring (not always appropriate), one can show that $FA \approx \asym J_1(j_{0,1}) \approx 0.52 \asym$, where $j_{0,1}$ is the first root of $J_0(x)$, while $J_0(x)$ and $J_1(x)$ are zeroth and first Bessel functions of the first kind. They also introduce a second metric, $IA \equiv A - 1$, where $A$ is the \cite{medeiros_asym_2022} asymmetry.  For 2017 M87* data they find $FA = 0.20-0.24$, implying $\asym \approx 0.38 - 0.46$, consistent with the $\asym=0.41$ value we find by fitting all visibility amplitudes and closure phases using Comrade. \cite{faggert_asym_2024} find that $FA$ depends strongly on inclination and is almost independent of spin at inclinations appropriate to M87*  (see their Figure 4), which differs significantly from our results. This is likely due to differences in the underlying disk model. 

\cite{saurabh_2025} perform a detailed study of asymmetry dependence on geometric parameters, including spin, in semi-analytic RIAF models, as well as in a subset of GRMHD simulations. They define asymmetry as the ratio of total imaged fluxes on both sides of black hole's rotation axis. The authors find that RIAF models tend to produce more symmetric images than GRMHD ones, possibly due to the simplifying assumptions on the magnetic field structure and fluid flow in RIAF models. Despite that, both RIAF and GRMHD models demonstrate asymmetry growth with $|\spin|$ and disfavor small negative $\spin$ models as too symmetric, which broadly agrees with our findings.

\subsection{Future EHT Observations}
\label{subsec:future_observations}

Future upgrades to EHT and future observations will improve our understanding of the ring asymmetry in M87* and Sgr A*.  In both sources, the emitting plasma is turbulent and samples a turbulent configuration about once per correlation time $\tau_{corr}$. For the asymmetry magnitude, the correlation time $\tau_{corr}(\asym) \approx 95\,t_g$ across the full model library. For MAD models, $\tau_{corr}(\asym) \approx 115\,t_g$. Thus for M87*, $\tau_{corr} \approx 100 \, t_g \approx  35$ days and for Sgr A* $\tau_{corr} \approx 30$ min. 

In Sgr A*, which we have not discussed yet, instantaneous $(u,v)$ coverage is poor compared to M87*.  Each turbulent configuration persists for only $\sim 30$ min, during which Earth's rotation causes the baselines to trace out only short arcs in the $(u,v)$ plane.  Thus, Sgr A*'s asymmetry is comparatively poorly constrained.  The addition of new antennas can improve coverage, allow better asymmetry estimates, and enable a model-dependent constraint on the inclination and spin of Sgr A*.  

In M87*, each turbulent configuration persists for $\sim 35$ days; during a night of observation, the source changes very little, but Earth's rotation causes the baselines to trace out long arcs in the $(u,v)$ plane (so-called Earth rotation aperture synthesis). 
As shown in Figure \ref{fig:uv_img}, this coverage permits accurate asymmetry measurements.  At present, there are three snapshots from uncorrelated turbulent configurations separated by a year, in 2017, 2018, and 2021.  The uncertainty in the measurement of the average asymmetry scales with the number of independent samples $N$ as $\sim \sigma/\sqrt{N}$, where $\sigma$ is the width of the ``true'' asymmetry distribution ($\sigma \simeq 0.2$, per Table \ref{tab:asym_gauss_fits}).  Existing data that has not been analyzed yet, from 2022–2024, will take $N$ from $3$ to $6$. We forecast that this will reduce uncertainty in the mean asymmetry by $\sim 1/\sqrt{2}$. 

ALMA has recently approved a time-lapse imaging campaign of M87* in 2026. If the asymmetry correlation times in M87* and models are consistent, then the main campaign will span $\sim2$ correlation times, while potential extensions would enable up to $\sim 4$ independent measurements.  Taken with 2025 observations and potential future observations in 2027+, EHT could achieve 12 independent samples by the end of the decade, reducing uncertainty in the mean asymmetry by a factor of $\sim 1/2$.

\begin{figure}
    \centering
    \includegraphics[width=0.8\textwidth]{ 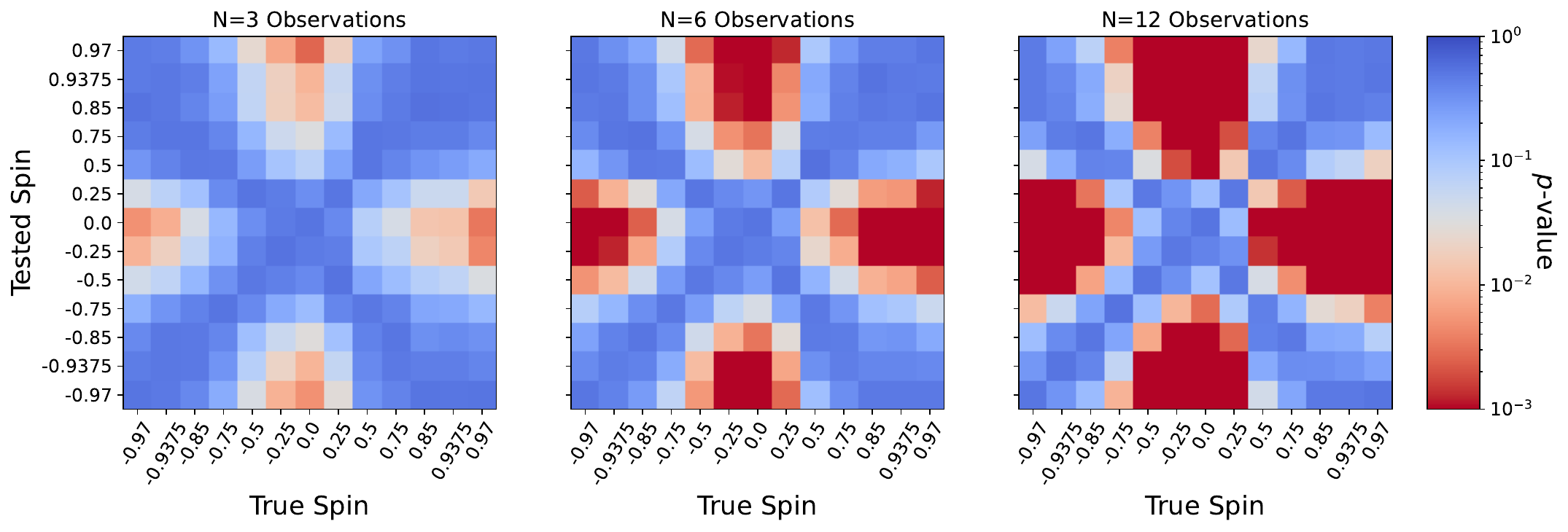}
    \caption{Effect of new observations on spin constraints. We repeatedly ``observe'' $N$ random asymmetry samples from the MAD model distribution for each ``true'' spin, then compare the mock observations to the asymmetry distribution for each ``tested'' spin using a KS test. 
    The resulting median probability $p$ of each tested spin is shown for $N = 3$ observations (left), $N = 6$ (center), and $N = 12$ (right). 
    High $p$ (blue) suggests the asymmetry distribution in the tested spin is a good match to the observations of the true spin; low $p$ (red) suggests we can rule out the tested spin.  As $N$ increases, we can correctly rule out an increasing fraction of the parameter space.
    }
    \label{fig:obs_diff_spins}
\end{figure}

To assess how future observations might constrain spin, we consider two scenarios: one showing the confidence with which spin zero models can be ruled out assuming future observations that are consistent with present constraints, and another showing how well the spin can be measured if the true asymmetry distribution in M87* is well described by one of our models.

For the first case, we assume the asymmetry is a (truncated) normal distribution with $\mu$ equal to the average of the three existing measurements, $\sim 0.55$, and the standard deviation $\simeq 0.2$, motivated by the typical standard deviation in the model library. We draw $N$ random samples (mock observations) from the truncated normal distribution, and perform a KS test comparing the mock observations to all spin zero models. For each $N$, we repeat this procedure many times.  We find the median $p$ for spin zero drops off as $p \sim e^{-N/1.6}$: the constraint improves exponentially as new observations are added. 

For the second case, we assume the correct model is in our library and ask how well future observations can measure the spin. The result is shown in Figure \ref{fig:obs_diff_spins}.  We fix the ``true'' spin of M87*, draw samples from the distribution for that spin reported in Table \ref{tab:asym_gauss_fits}, and test the $N$ simulated observations against the model distribution at each spin.  We again repeat this procedure many times and take the median $p$. The resulting $p$ for different M87* spins and observational samples $N$ are shown in Figure \ref{fig:obs_diff_spins}. The $p$-value recovered for the true spin lies on the ``True Spin = Tested Spin'' diagonal.  If M87* is highly spinning, we will be able to strongly exclude zero-spin models by $N=6$. If on the other hand M87* has low spin, we will also be able to strongly exclude high spin by $N=6$.

\subsection{Degeneracy with Inclination}\label{sec:incl}

So far, we have only considered $i=17^\circ$ and $163^\circ$. As earlier works such as \cite{medeiros_asym_2022} have noted, however, the asymmetry also varies with inclination. When face-on at $i=0^\circ$ or $180^\circ$, we would expect $\asym=0$; when edge-on, we would expect $\asym$ to reach its maximum.  To what degree does the uncertainty in M87*'s estimated inclination \citep[with standard deviation $\approx3^\circ$--$4^\circ$, per][]{Mertens_jet_2016, walker_jet_2018} contaminate our spin constraints?

We assess the uncertainty introduced by inclination error in a small sample of models, changing the inclination by $\pm 5^\circ$ from the inclination used in our standard model series.  We consider four MAD models, with $\rhigh=40$, $R_{\textrm{low}}=1$, and $\spin=\pm 0.5$.  We sample 30 images from each model separated by $> 50\, t_g$. In the prograde model, we find a shift of $i\pm5^\circ$ corresponds to a shift in asymmetry of $\asym\pm 0.1$. In the retrograde model, we find a shift of $i\pm5^\circ$ corresponds to a shift in asymmetry of $\asym \pm 0.04$.  The asymmetry uncertainty due to inclination error $\sigma_{i}$ will begin to dominate when $\sigma/\sqrt{N} < \sigma_{i}$ for $N$ independent samples and asymmetry distribution standard deviation $\sigma$. This occurs when $N \gtrsim 4$ for the retrograde models, and $N \gtrsim 25$ for the prograde models.  

\subsection{Model Uncertainty}\label{sec:uncert}

Our asymmetry distributions are model-dependent.  What are the sources of model uncertainty?

Two uncertainties (magnetization and adiabatic index) are discussed in detail in Appendix \ref{sec:appendix}, since they can be evaluated from existing simulation data.  To summarize: across v3 and v5 MAD and SANE models, 
the asymmetry is minimized at zero or at slightly negative spin, although the exact relationship between asymmetry and spin shows some variation.  Additional sources of model uncertainty include the following:

(1) Electron energization and thermodynamics.  Our models assume the electron distribution function is thermal with $T_e(\beta, T_i; \rhigh, \rlow)$, $T_e \ll T_i$, and with an isotropic $T_e $ relative to the local magnetic field. 
Better models would evolve a separate electron internal energy equation. This may produce better agreement of the models with Sgr A* variability data \citep{salas_2024}, although it requires modeling electron heating (an unsettled issue in the theory of collisionless plasmas), electron cooling  (particularly important in M87*), and acceleration of electrons into a nonthermal tail.  

GRMHD models with the nonthermal tail in the electron distribution function produce remarkably similar brightness asymmetries to their thermal counterparts \citepalias[see][ Figure 3]{EHTC_2025_1}. One set of GRMHD models that evolve a separate electron internal energy
shows similar trends in asymmetry, but has a decrease in the overall magnitude by $\sim 33\%$ after accounting for a different asymmetry definition, which should produce values $2\times$ smaller than ours if the data corresponds to an $m=1$ $m$-ring \citep{Chael_2025}. If the decrease is due to the differing models and not the differing asymmetry definition, then the smaller asymmetry would allow us to more confidently rule out zero spin.

Electron anisotropy can also affect asymmetry \citep{Galishnikova_23_anisotropy}. Anisotropy perpendicular to the magnetic field ($T_\perp > T_\parallel$, as in the mirror and whistler limits) produces increased asymmetry in the image and is disfavored based on the observed circular polarization fraction (\citealp{ginzburg_1965, Galishnikova_23_anisotropy}; \citetalias{EHTC_2023_9}).
Anisotropy parallel to the magnetic field ($T_\parallel > T_\perp$, as in the firehose limit) would decrease the observed asymmetry, thus increasing the confidence with which we could rule out zero spin. 

(2) Collisionless dynamics.  Our models treat the accreting plasma as an ideal fluid, but the plasma is known to be Coulomb-collisionless.\footnote{Ideal here refers to the governing equations.  The numerical implementation, which is an ILES or Implicit Large Eddy Simulation, still incorporates dissipation due to truncation error at the grid scale.  See \cite{grete_2023} for a discussion of ILES in compressible MHD.}  The accuracy of this approximation is not yet understood.  Although Coulomb scattering is too weak to fluidize the plasma, wave-particle scattering may be strong enough \citep[e.g.][]{kunz_2014}.  One particular source of concern is that, at least for some model problems, the reconnection rate is an order of magnitude slower in MHD than in kinetic theory \citep[e.g.][]{ripperda_2020}.  Models that incorporate nonideal corrections (viscosity and heat conduction) to an ideal fluid show behavior that is remarkably similar to ideal models \citep{foucart_2017, Dhruv_nonideal_2025}.  

(3) Boundary and initial conditions.  The models are initialized with a magnetized torus of limited radial extent \citep{Fishbone_moncrief_torus, wong_patoka_2022}. 
The variation of model outcomes with respect to changes in this initial condition and boundary conditions have not yet been fully assessed.  It is known that in models of Sgr A*, a limited set of self-consistent wind-fed models \citep{ressler_ab_2020} produce images that are similar to images produced with our initial conditions used here \citepalias{EHTC_2022_5}.  

(4) The ``fast light'' approximation. The fast light approximation used here performs the radiative transfer calculation for a snapshot on a single time slice.  A more precise treatment (``slow light'') would trace the photon trajectories through multiple time slices, i.e. through the fluid as it evolves, since the light crossing time is only slightly less than the dynamical time close to the horizon. The fast light approximation is used because it is cheap and simple. We have compared the asymmetry distribution in fast light and slow light versions of a test model and find that they are nearly identical. The asymmetry had a mean and standard deviation of $0.70\pm0.18$ in the fast light model, while it was $0.70\pm0.19$ in the slow light model.

Finally, there other sources of uncertainty such as those related to numerical treatment of low-density regions in the flow.  Ultimately the models will be tested by comparison with observations.

\subsection{Testing the Blandford–Znajek Jet-Launching Mechanism} 
\label{subsec:bz}
The Blandford–Znajek (BZ) mechanism \citep{blandford_znajek} is commonly considered the favored mechanism for jet-launching. BZ predicts (1) that there is a black hole at the base of M87*'s jet, (2) that the jet connects to the black hole, (3) that the black hole is spinning, and (4) that the jet power is a function of black hole spin.  The presence of a black hole is unambiguously confirmed by $230$ GHz VLBI imaging, which shows the expected ring  \citepalias{EHTC_2019_1}. $86$ GHz VLBI imaging shows a connection between the black hole and jet \citep{lu_2023}. Our analysis suggests that the black hole is spinning---consistent with BZ. Estimates of M87*'s jet power typically favor higher spins assuming BZ 
(\citealp{Nemmen_2019_jetpower}; \citetalias{EHTC_2019_5}), and these higher spins are among the values favored by asymmetry.
In Section \ref{subsec:future_observations}, we show future observations may provide improved spin constraints. Future EHT or space-VLBI observations could test BZ by checking for continued consistency between estimates of spin and jet power.  
Finally, future observations may also provide a direct probe of the electromagnetic energy extraction direction using polarization measurements, which could also test BZ \citep{chael_2023}. 

\section{Conclusion} \label{sec:conclusion}

In this paper, we have presented tentative evidence for a spinning M87* black hole based on the asymmetry of 2017, 2018, and 2021 EHT images.  The evidence is tentative because it relies on (1) the relevance of the model library itself, which is known to have missing physical ingredients and internal inconsistencies, discussed in Section \ref{sec:uncert}; (2) a restriction of the model library to magnetically arrested disk (MAD) models, which are nevertheless preferred by EHT data; and (3) only three independent observations.  With these assumptions, zero-spin models are disfavored.

The spin-ring asymmetry correlation is clearer for MAD models than for SANE models (see Appendix \ref{sec:appendix}), and MAD models are mildly favored by existing model--data comparisons. SANEs are more likely to fail the jet power constraint \citepalias{EHTC_2019_5}. SANEs are also either too weakly polarized, or have weak magnetic field lines that become too azimuthal as they are dragged by the fluid, producing EVPAs that are too radial to agree with the observed orientations \citepalias{EHTC_2021_8}.

At present, there are three independent asymmetry measurements for M87*.  This paper highlights the value of new data for constraining the spin of M87*. In Section \ref{subsec:future_observations}, we have calculated the median probability that the observations are drawn from an $\spin = 0$ model asymmetry distribution to scale as $\exp(-N/1.6)$, where $N$ is the number of independent asymmetry measurements.   

Our analysis, based on image asymmetry alone, suggests that M87* may have  $|\spin| \gtrsim 0.2$, as can be seen in Figure \ref{fig:p_values}. In particular, $\spin = 0$ and $-0.25$ models are marginally disfavored at $p\approx0.026$ and $0.049$ respectively. Models with $\spin = 0.25$ are currently still plausible at $p \approx 0.14$. The EHT could strengthen its spin constraints with an increased sample size, e.g., using future time-lapse imaging of M87*.  This is consistent with previous arguments based on jet power \citepalias{EHTC_2019_5}: low-spin models that match the compact millimeter flux of M87*'s accretion disk have jet power far below existing estimates.  Our argument here is independent of the jet power argument, and does not rely on highly uncertain jet power estimates.

Our spin constraint is also consistent with spin distributions obtained in \cite{janssen_2025}, which uses machine learning models trained on GRMHD simulations to interpret polarized data. Our analysis does not use M87* polarization data. The EHT could also strengthen its spin constraints by incorporating other observables, such as the aforementioned jet power, polarization structure of the ring \citep{palumbo_2020, chael_2023}, and pattern speed \citep{Conroy_2023}. 

The asymmetry argument presented here, the jet power argument \citepalias{EHTC_2019_5}, and the outcome of GRMHD simulations \citep[e.g.][]{mckinney_2004} are all consistent with the notion that the M87* jet is powered by the BZ mechanism. We outline several methods to test predictions of the BZ mechanism with future observations.

A future space-based VLBI antenna, such as the proposed NASA BHEX mission \citep{bhex_2024}, would enable us to resolve near-horizon rings in sources other than M87* and Sgr A* \citep{zhang_targets_2024}.  Our work suggests that it may be possible to estimate the magnitude of spin in these sources based on asymmetry alone, if a sufficiently good estimate for the inclination is available. Ring diameter and brightness asymmetry are perhaps the easiest features to measure in EHT images, thus enabling future analysis to provide model-dependent constraints on mass and spin across a variety of sources.

\appendix
\section{Other Model Sets}
\label{sec:appendix}
\subsection{v3 Illinois GRMHD Library}
In this study, we rely on the v5 version of the \patoka model library. This version has significant differences from version v3. The most notable physical difference is that in v5, the plasma adiabatic index is $\gamma = \frac{5}{3}$, compared to $\gamma = \frac{4}{3}$ in v3. Models in v5 were run longer, at higher resolution, with a slight change in $\rhigh$ (v3 has $\rhigh=1$ instead of $80$), and included more spins (v3 only modeled $5$ spin values $\spin \in \{-0.94, -0.5, 0, 0.5, 0.94\}$, compared to $13$ spin values modeled in v5). In addition, v3 utilized an older \iharm code \citep{prather_iharm_2021} to evolve the fluid, compared to the newer \kharma code used in v5.

The effect of each of these differences has yet to be fully assessed. However, the statistical characteristics of synthetic images can vary significantly between v3 and v5. The adiabatic index of $\gamma = 5/3$ in v5 produces higher temperatures in the emitting region than the lower adiabatic index of $\gamma = 4/3$ in v3. The higher adiabatic index of the v5 library analyzed in the body of this text is considered more accurate \citep{Gammie_2025}.

In Figure \ref{fig:model_comparison}, we show the asymmetry dependence for v3 and v5 models, again marginalizing across the radiative transfer parameters $\rhigh$ and $\rlow$ as above. Despite differences, the v3 model library generally produces similar trends to v5. Both are maximized at $|\spin| \approx 1$ and minimized at $|\spin| \ll 1$. One interesting feature of the v3 models is that the minimum in median asymmetry occurs at $\spin = -0.5$, which is different from the minimum at $\spin = 0$ in v5.  The reason for this difference is likely due to different velocity profiles in the emission region. The outdated, lower adiabatic index is expected to enable faster rotation \citep[See the Appendix of][]{Conroy_2025}. Then for v3 $\spin = -0.5$, Doppler beaming may contribute brighter emission on the side of approaching disk and receding spin, counteracting the effects of the moderate retrograde spin to create a more symmetric ring.
However, we note the v3 library has no spins sampled between $\spin = 0$ and $-0.5$, so it is possible a denser sampling of spins would produce a closer minimum.

\subsection{SANE models}
\begin{figure}
    \centering
    \includegraphics[width=0.6\textwidth]{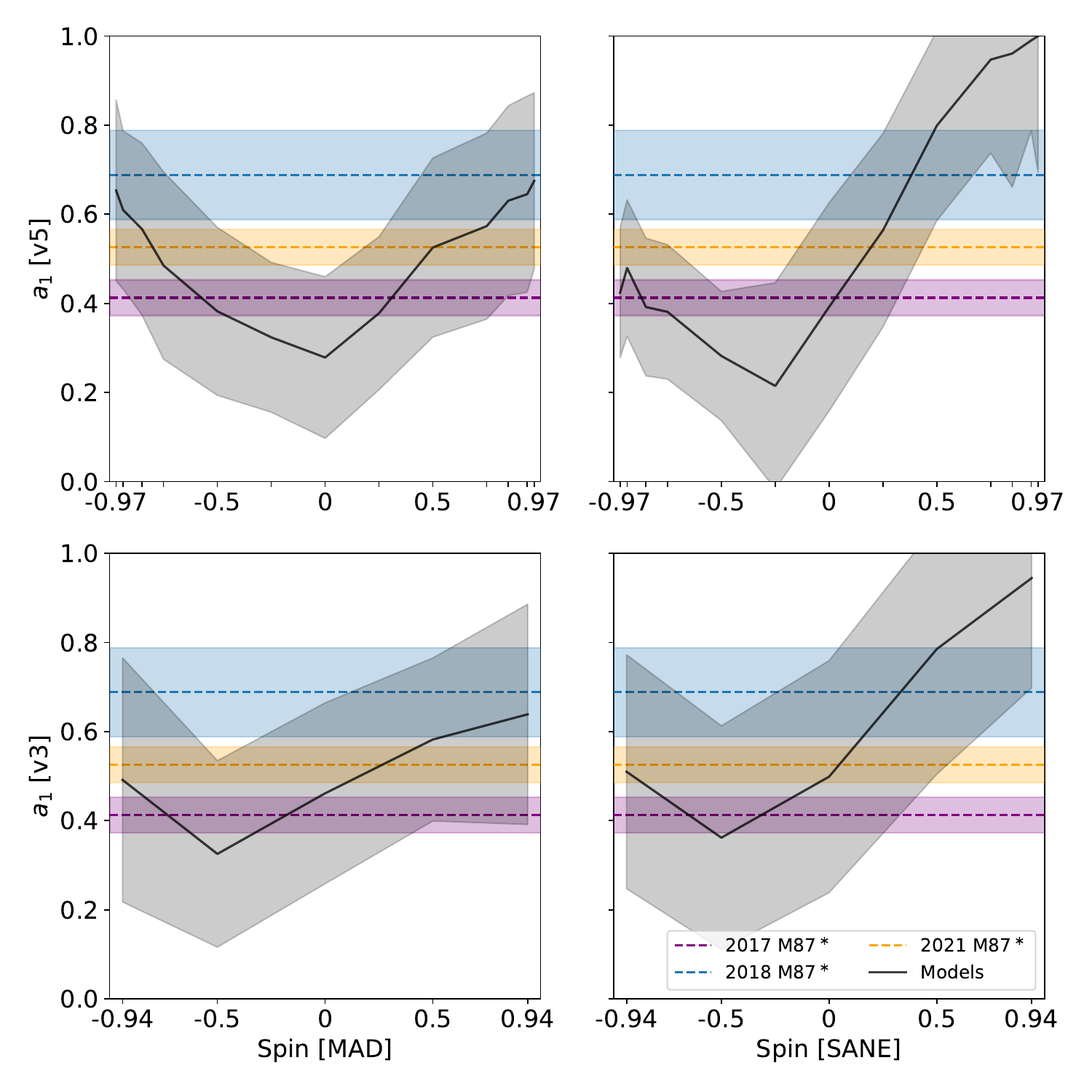}
    \caption{Asymmetry curves with $1\sigma$ ranges, as in Figure \ref{fig:spin_distr}, for v5 MAD (top left), v5 SANE (top right), v3 MAD (bottom left), v3 SANE (bottom right) models. The black lines show the modes $\mu$ of truncated Gaussians fitted to modeled asymmetry distributions at each spin. The gray bands show the $1\sigma$  variation of the distributions. Observed asymmetries in M87* are marked with dashed lines.  The colored bands show $1\sigma$ uncertainties. X-axis tick marks show the spins sampled in each set of models.}
    \label{fig:model_comparison}
\end{figure}

We also show the brightness asymmetry in SANEs in Figure \ref{fig:model_comparison}. In both model libraries, the mean asymmetry is larger for prograde SANEs than retrograde SANEs, in contrast to MADs where the distribution is more symmetrical. In SANEs, the rotation curve is nearly Keplerian, while MADs have strongly sub-Keplerian rotation curves \citep{Conroy_2023, Dhruv_grmhd_survey_2024}. Thus, Doppler beaming plays a larger role in SANEs relative to lensing, increasing the asymmetry for prograde models.\footnote{Note that asymmetry measurements for v5 SANEs at $\spin=0.85$ and $0.97$ were calculated over a shorter duration ($t=20,000 $ to $30,000 \, t_g$ rather than $20,000 $ to $50,000 \, t_g$), with a slightly higher cadence (every $100\,t_g$ rather than every $125\,t_g$), using a reduced image resolution. These differences may account for the wider $1\sigma$ region for those models.}
This makes SANEs better agree with observed asymmetries of M87* at more modest positive spin values. For retrograde models, Doppler beaming and lensing compete (along with other effects) to produce asymmetry in opposite hemispheres. 

\section{$m$-ring Truncation, Asymmetry Bounds, and $(u,v)$ Coverage}
\label{subsec:appendix_bounds_coverage}

For the bulk of this analysis, we used an $m$-ring model that was truncated at $m=4$, bound by asymmetry $a_m\in[0,1)$, and fit in the $(u,v)$ domain assuming 2018 $(u,v)$ coverage. In this section, we show that changes to these specific methodology choices do not affect our spin constraints.

Our $m=4$ $m$-ring had zero ellipticity and was allowed a single nuisance Gaussian, which models diffuse off-ring emission. Geometric ring models are well-motivated for EHT sources (e.g., \citetalias{EHTC_2019_4}; \citealp{tiede_comrade_2022}; \citealp{tiede_vida_2022}), especially $m$-rings with $m\in\{2, 3, 4\}$. The addition of a nuisance Gaussian improves the fit of the model \citepalias[Figure 14]{EHTC_2024_1}.  Here, for simplicity, we considered a single Comrade template.  Would different choices of geometric model template, parameter bounds, or simulated $(u,v)$ coverage change our results?

\begin{figure}[h]
    \centering
    \includegraphics[width=0.5\textwidth]{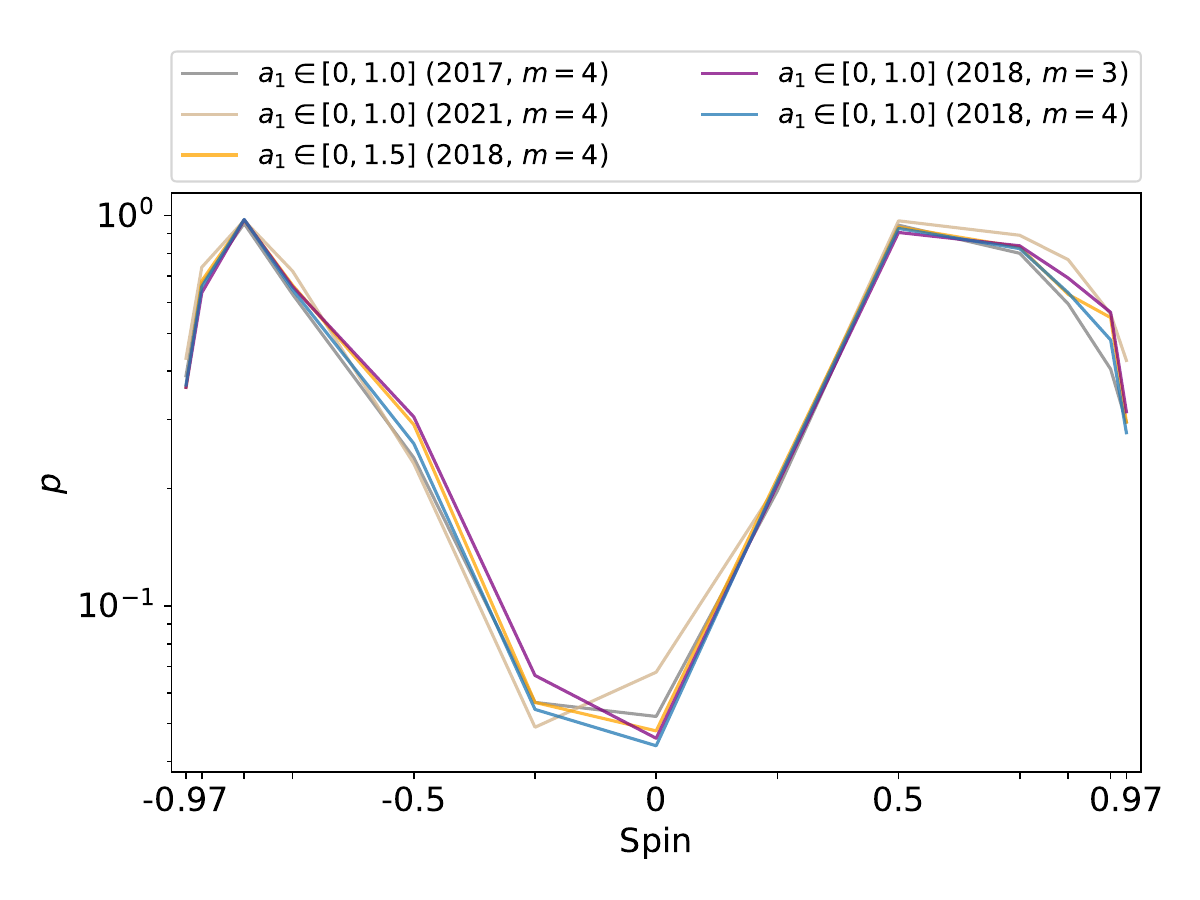}
    \caption{Same as Figure \ref{fig:p_values}, but now showing the probability $p$ across spin for the two-sample KS test using different modeling methodologies. We replot the curve of Figure \ref{fig:p_values} (blue), adjust the assumed $(u,v)$ coverage from 2018 to 2017 (gray) and 2021 (tan), adjust the Comrade template to allow a maximum asymmetry of $a_{1 \, \textrm{max}}=1.5$ (orange), and adjust the Comrade template to use an $m=3$ $m$-ring (purple). The results are largely consistent, producing lower $p$-values for zero-spin models.}
    \label{fig:other_modeling}
\end{figure}

Figure \ref{fig:other_modeling} shows two-sample KS test as done in Figure \ref{fig:p_values}, this time while varying assumed $(u,v)$ coverage from 2017 to 2021, the assumed parameter bounds from $a_{1 \,\textrm{max}}=1$ to $1.5$, and the geometric model template from an $m=4$ $m$-ring to an $m=3$ one.  This analysis was performed at a reduced cadence of one frame every $500 \, t_g$, leaving 60 samples for each unique model, and 480 for each unique spin.  The changes in asymmetry bounds or assumed $(u,v)$ coverage have a negligible effect on $p$-values. 

A small subset of MAD model frames ($2.6\%$ of all MAD frames, $0.3\%$ of MAD $|\spin|\leq0.5$ frames) have asymmetry near the boundary ($\asym >0.99$, as is visible in Figure \ref{fig:uv_img}). These are most common in high-spin models with a large brightness peak. By visual inspection of GRMHD frames, $\asym \simeq 1$ occurs when the dim side of the ring is particularly dim, or when the ring appears obscured with a large amount of extended emission. In this case, setting a higher asymmetry bound allows us to capture the tail of asymmetry distribution in $|\spin| \sim1$ models.  Although $\asym > 1$ would appear to permit negative intensity on the faint side of the ring, the optimizer prevents this by aligning the phase of higher order $m$-rings and the nuisance Gaussian. Thus, recovering such large asymmetry requires accurate recovery of the higher-order $m$-ring harmonics. 

Rather than shifting the bound of the truncated Gaussian, one might also consider fitting a clipped Gaussian distribution, where the values outside of $\asym = 1$ are reflected across the threshold. We tried fitting a clipped Gaussian and found this had little effect on the recovered $\mu$ for each model. In any event, current measurements of asymmetry observations are far from the $\asym =1$ boundary, and raising the boundary has negligible effect on $p$-values. 

Finally, we tried fitting an $m=3$ ring to observations and compared to the distribution found in models at each spin and again found largely consistent results. Zero-spin models produce a low $p$-value. We conclude that our spin constraints are relatively insensitive to details of the geometric model. As suggested by Figure \ref{fig:obs_diff_spins}, if the true spin magnitude of M87* is far enough from zero, we will be able to more confidently rule out zero-spin models using future observations.

\section{Asymmetry Fits from GRMHD Library}
\label{subsec:appendix_fits}

Table \ref{tab:asym_gauss_fits_all} provides the mode  $\mu$ and the standard deviation $\sigma$ of the truncated Gaussian fits to each model in the v5 M87* library. The model parameters are summarized in Table \ref{tab:model_params}. As in Table \ref{tab:asym_gauss_fits}, we filter out the small percentage of frames that have an asymmetry equal to the bound when fitting the truncated normal parameters. A small fraction of high-spin SANEs are poorly fit, as the asymmetry boundary truncates a significant portion of the asymmetry distribution.
Nevertheless, the chosen bounds generally improve fitting accuracy, and have negligible effects on observational constraints. 
See Appendix \ref{sec:appendix} and \ref{subsec:appendix_bounds_coverage} for discussion.

\begin{deluxetable}{ccccccc}
    \label{tab:asym_gauss_fits_all}
    \tablecaption{Gaussian Fits to Asymmetry for All Models}
    \tablewidth{\textwidth}
    \tablehead{
    \colhead{MAD/SANE} & \colhead{$\spin$} & \colhead{$i \, [^\circ]$} &
    \colhead{$\rhigh$} & \colhead{$\rlow$} &
    \colhead{$\mu$} & \colhead{$\sigma$}
    }
    \startdata
    MAD & $ -0.97 $ & $ 17 $ & $ 10 $ & $ 1 $ & $ 0.612 $ & $ 0.192 $ \\
    MAD & $ -0.97 $ & $ 17 $ & $ 10 $ & $ 10 $ & $ 0.622 $ & $ 0.214 $ \\
    MAD & $ -0.97 $ & $ 17 $ & $ 40 $ & $ 1 $ & $ 0.671 $ & $ 0.194 $ \\
    MAD & $ -0.97 $ & $ 17 $ & $ 40 $ & $ 10 $ & $ 0.663 $ & $ 0.214 $ \\
    MAD & $ -0.97 $ & $ 17 $ & $ 80 $ & $ 1 $ & $ 0.671 $ & $ 0.189 $ \\
    MAD & $ -0.97 $ & $ 17 $ & $ 80 $ & $ 10 $ & $ 0.656 $ & $ 0.201 $ \\
    MAD & $ -0.97 $ & $ 17 $ & $ 160 $ & $ 1 $ & $ 0.674 $ & $ 0.207 $ \\
    MAD & $ -0.97 $ & $ 17 $ & $ 160 $ & $ 10 $ & $ 0.661 $ & $ 0.207 $ \\
    \enddata
    \tablecomments{Truncated Gaussian fits to asymmetry distributions of individual GRMHD models in v5 library. Each model is specified by its magnetic field mode (MAD/SANE), spin $\spin$, observer inclination $i$, and electron temperature parameters $\rhigh$ and $\rlow$. The fitted mode $\mu$ and standard deviation $\sigma$ are shown. Only a portion of this table is shown here to demonstrate its form and content. (This table is available in its entirety in machine-readable form in the online article.)}
\end{deluxetable}

\acknowledgements

We thank the anonymous referee for their thorough comments which greatly improved this work.
This work was supported by NSF grants AST 17-16327 (horizon), OISE 17-43747, and AST  20-34306. This material is based upon work supported by the National Science Foundation under Grant No. PHY-2244433. 
N.C. is supported by the NASA Future Investigators in NASA Earth and Space Science and Technology (FINESST) program. This material is based upon work supported by the National Aeronautics and Space Administration under Grant No. 80NSSC24K1475 issued through the Science Mission Directorate.
This research used resources of the Oak Ridge Leadership 
Computing Facility at the Oak Ridge National Laboratory, which is 
supported by the Office of Science of the U.S. Department of Energy 
under Contract No. DE-AC05-00OR22725.  This research used resources of
the Argonne Leadership Computing Facility, which is a DOE Office of 
Science User Facility supported under Contract DE-AC02-06CH11357.  
This research was done using services provided by the OSG Consortium, which is supported by the National Science Foundation awards \#2030508 and \#1836650.
  This research is part of the Delta research computing project, which 
is supported by the National Science Foundation (award OCI 2005572), and the State of Illinois. Delta is a joint effort of the University of Illinois at Urbana-Champaign and its National Center for Supercomputing Applications. We are particularly grateful to the Argonne Leadership Computing Facility for providing storage space on the Eagle system that was critical for enabling this work.  This work was initiated in part at the Aspen Center for Physics, which is supported by National Science Foundation grant PHY-2210452. 

The Event Horizon Telescope Collaboration thanks the following
organizations and programs: the Academia Sinica; the Research Council of Finland (project 362572); the Agencia Nacional de Investigaci\'{o}n 
y Desarrollo (ANID), Chile via NCN$19\_058$ (TITANs), Fondecyt 1221421 and BASAL FB210003; the Alexander
von Humboldt Stiftung (including the Feodor Lynen Fellowship); an Alfred P. Sloan Research Fellowship;
Allegro, the European ALMA Regional Centre node in the Netherlands, the NL astronomy
research network NOVA and the astronomy institutes of the University of Amsterdam, Leiden University, and Radboud University;
the ALMA North America Development Fund; the Astrophysics and High Energy Physics programme by MCIN (with funding from European Union NextGenerationEU, PRTR-C17I1); the Black Hole Initiative, which is funded by grants from the John Templeton Foundation (60477, 61497, 62286) and the Gordon and Betty Moore Foundation (Grant GBMF-8273) - although the opinions expressed in this work are those of the author and do not necessarily reflect the views of these Foundations; 
the Brinson Foundation; the Canada Research Chairs (CRC) program; Chandra DD7-18089X and TM6-17006X; the China Scholarship
Council; the China Postdoctoral Science Foundation fellowships (2020M671266, 2022M712084); ANID through Fondecyt Postdoctorado (project 3250762); Conicyt through Fondecyt Postdoctorado (project 3220195); Consejo Nacional de Humanidades, Ciencia y Tecnología (CONAHCYT, Mexico, projects U0004-246083, U0004-259839, F0003-272050, M0037-279006, F0003-281692, 104497, 275201, 263356, CBF2023-2024-1102, 257435); the Colfuturo Scholarship; the Consejo Superior de Investigaciones 
Cient\'{i}ficas (grant 2019AEP112);
the Delaney Family via the Delaney Family John A.
Wheeler Chair at Perimeter Institute; Dirección General de Asuntos del Personal Académico-Universidad Nacional Autónoma de México (DGAPA-UNAM, projects IN112820 and IN108324); the Dutch Research Council (NWO) for the VICI award (grant 639.043.513), the grant OCENW.KLEIN.113, and the Dutch Black Hole Consortium (with project No. NWA 1292.19.202) of the research programme the National Science Agenda; the Dutch National Supercomputers, Cartesius and Snellius  (NWO grant 2021.013); 
the EACOA Fellowship awarded by the East Asia Core
Observatories Association, which consists of the Academia Sinica Institute of Astronomy and Astrophysics, the National Astronomical Observatory of Japan, Center for Astronomical Mega-Science,
Chinese Academy of Sciences, and the Korea Astronomy and Space Science Institute; 
the European Research Council (ERC) Synergy Grant ``BlackHoleCam: Imaging the Event Horizon of Black Holes'' (grant 610058) and Synergy Grant ``BlackHolistic:  Colour Movies of Black Holes:
Understanding Black Hole Astrophysics from the Event Horizon to Galactic Scales'' (grant 10107164); 
the European Union Horizon 2020
research and innovation programme under grant agreements
RadioNet (No. 730562), 
M2FINDERS (No. 101018682); the European Research Council for advanced grant ``JETSET: Launching, propagation and 
emission of relativistic jets from binary mergers and across mass scales'' (grant No. 884631); the European Horizon Europe staff exchange (SE) programme HORIZON-MSCA-2021-SE-01 grant NewFunFiCO (No. 10108625); the Horizon ERC Grants 2021 programme under grant agreement No. 101040021; the FAPESP (Funda\c{c}\~ao de Amparo \'a Pesquisa do Estado de S\~ao Paulo) under grant 2021/01183-8; the Fondes de Recherche Nature et Technologies (FRQNT); the Fondo CAS-ANID folio CAS220010; the Generalitat Valenciana (grants APOSTD/2018/177 and  ASFAE/2022/018) and
GenT Program (project CIDEGENT/2018/021); the Gordon and Betty Moore Foundation (GBMF-3561, GBMF-5278, GBMF-10423);   
the Institute for Advanced Study; the ICSC – Centro Nazionale di Ricerca in High Performance Computing, Big Data and Quantum Computing, funded by European Union – NextGenerationEU; the Istituto Nazionale di Fisica
Nucleare (INFN) sezione di Napoli, iniziative specifiche
TEONGRAV; 
the International Max Planck Research
School for Astronomy and Astrophysics at the
Universities of Bonn and Cologne; the Italian Ministry of University and Research (MUR)– Project CUP F53D23001260001, funded by the European Union – NextGenerationEU; 
Deutsche Forschungsgemeinschaft (DFG) research grant ``Jet physics on horizon scales and beyond'' (grant No. 443220636) and DFG research grant 443220636;
Joint Columbia/Flatiron Postdoctoral Fellowship (research at the Flatiron Institute is supported by the Simons Foundation); 
the Japan Ministry of Education, Culture, Sports, Science and Technology (MEXT; grant JPMXP1020200109); 
the Japan Society for the Promotion of Science (JSPS) Grant-in-Aid for JSPS
Research Fellowship (JP17J08829); the Joint Institute for Computational Fundamental Science, Japan; the Key Research
Program of Frontier Sciences, Chinese Academy of
Sciences (CAS, grants QYZDJ-SSW-SLH057, QYZDJSSW-SYS008, ZDBS-LY-SLH011); 
the Leverhulme Trust Early Career Research
Fellowship; the Max-Planck-Gesellschaft (MPG);
the Max Planck Partner Group of the MPG and the
CAS; the MEXT/JSPS KAKENHI (grants 18KK0090, JP21H01137,
JP18H03721, JP18K13594, 18K03709, JP19K14761, 18H01245, 25120007, 19H01943, 21H01137, 21H04488, 22H00157, 23K03453); the MICINN Research Projects PID2019-108995GB-C22, PID2022-140888NB-C22; the MIT International Science
and Technology Initiatives (MISTI) Funds; 
the Ministry of Science and Technology (MOST) of Taiwan (103-2119-M-001-010-MY2, 105-2112-M-001-025-MY3, 105-2119-M-001-042, 106-2112-M-001-011, 106-2119-M-001-013, 106-2119-M-001-027, 106-2923-M-001-005, 107-2119-M-001-017, 107-2119-M-001-020, 107-2119-M-001-041, 107-2119-M-110-005, 107-2923-M-001-009, 108-2112-M-001-048, 108-2112-M-001-051, 108-2923-M-001-002, 109-2112-M-001-025, 109-2124-M-001-005, 109-2923-M-001-001, 
110-2112-M-001-033, 110-2124-M-001-007 and 110-2923-M-001-001); the National Science and Technology Council (NSTC) of Taiwan
(111-2124-M-001-005, 112-2124-M-001-014,  112-2112-M-003-010-MY3, and 113-2124-M-001-008);
the Ministry of Education (MoE) of Taiwan Yushan Young Scholar Program;
the Physics Division, National Center for Theoretical Sciences of Taiwan;
the National Aeronautics and
Space Administration (NASA, Fermi Guest Investigator
grant 
80NSSC23K1508, NASA Astrophysics Theory Program grant 80NSSC20K0527, NASA NuSTAR award 
80NSSC20K0645); NASA Hubble Fellowship Program Einstein Fellowship;
NASA Hubble Fellowship 
grants HST-HF2-51431.001-A, HST-HF2-51482.001-A, HST-HF2-51539.001-A, HST-HF2-51552.001A awarded 
by the Space Telescope Science Institute, which is operated by the Association of Universities for 
Research in Astronomy, Inc., for NASA, under contract NAS5-26555; 
the National Institute of Natural Sciences (NINS) of Japan; the National
Key Research and Development Program of China
(grant 2016YFA0400704, 2017YFA0402703, 2016YFA0400702); the National Science and Technology Council (NSTC, grants NSTC 111-2112-M-001 -041, NSTC 111-2124-M-001-005, NSTC 112-2124-M-001-014); the US National
Science Foundation (NSF, grants AST-0096454,
AST-0352953, AST-0521233, AST-0705062, AST-0905844, AST-0922984, AST-1126433, OIA-1126433, AST-1140030,
DGE-1144085, AST-1207704, AST-1207730, AST-1207752, MRI-1228509, OPP-1248097, AST-1310896, AST-1440254, 
AST-1555365, AST-1614868, AST-1615796, AST-1715061, AST-1716327,  AST-1726637, 
OISE-1743747, AST-1743747, AST-1816420, AST-1935980, AST-1952099, AST-2034306,  AST-2205908, AST-2307887, AST-2407810); 
NSF Astronomy and Astrophysics Postdoctoral Fellowship (AST-1903847); 
the Natural Science Foundation of China (grants 11650110427, 10625314, 11721303, 11725312, 11873028, 11933007, 11991052, 11991053, 12192220, 12192223, 12273022, 12325302, 12303021); 
the Natural Sciences and Engineering Research Council of
Canada (NSERC); 
the National Research Foundation of Korea (the Global PhD Fellowship Grant: grants NRF-2015H1A2A1033752; the Korea Research Fellowship Program: NRF-2015H1D3A1066561; Brain Pool Program: RS-2024-00407499;  Basic Research Support Grant 2019R1F1A1059721, 2021R1A6A3A01086420, 2022R1C1C1005255, RS-2022-NR071771, 2022R1F1A1075115); the POSCO Science Fellowship of the POSCO TJ Park Foundation; NOIRLab, which is managed by the Association of Universities for Research in Astronomy (AURA) under a cooperative agreement with the National Science Foundation; 
Onsala Space Observatory (OSO) national infrastructure, for the provisioning
of its facilities/observational support (OSO receives funding through the Swedish Research Council under grant 2017-00648);  the Perimeter Institute for Theoretical Physics (research at Perimeter Institute is supported by the Government of Canada through the Department of Innovation, Science and Economic Development and by the Province of Ontario through the Ministry of Research, Innovation and Science); the Portuguese Foundation for Science and Technology (FCT) grants (Individual CEEC program – 5th edition, CIDMA
through the FCT Multi-Annual Financing Program for R\&D Units UID/04106, CERN/FIS-PAR/0024/2021, 2022.04560.PTDC); the Princeton Gravity Initiative; the Spanish Ministerio de Ciencia, Innovaci\'{o}n  y Universidades (grants PID2022-140888NB-C21, PID2022-140888NB-C22, PID2023-147883NB-C21, RYC2023-042988-I); the Severo Ochoa grant CEX2021-001131-S funded by MICIU/AEI/10.13039/501100011033; The European Union’s Horizon Europe research and innovation program under grant agreement No. 101093934 (RADIOBLOCKS); The European Union “NextGenerationEU”, the Recovery, Transformation and Resilience Plan, the CUII of the Andalusian Regional Government and the Spanish CSIC through grant AST22\_00001\_Subproject\_10; ``la Caixa'' Foundation (ID 100010434) through fellowship codes LCF/BQ/DI22/11940027 and LCF/BQ/DI22/11940030; 
the University of Pretoria for financial aid in the provision of the new 
Cluster Server nodes and SuperMicro (USA) for a SEEDING GRANT approved toward these 
nodes in 2020; the Shanghai Municipality orientation program of basic research for international scientists (grant no. 22JC1410600); 
the Shanghai Pilot Program for Basic Research, Chinese Academy of Science, 
Shanghai Branch (JCYJ-SHFY-2021-013); the Simons Foundation (grant 00001470); the Spanish Ministry for Science and Innovation grant CEX2021-001131-S funded by MCIN/AEI/10.13039/501100011033; the Spinoza Prize SPI 78-409; the South African Research Chairs Initiative, through the 
South African Radio Astronomy Observatory (SARAO, grant ID 77948),  which is a facility of the National 
Research Foundation (NRF), an agency of the Department of Science and Innovation (DSI) of South Africa; the Swedish Research Council (VR); the Taplin Fellowship; the Toray Science Foundation; the UK Science and Technology Facilities Council (grant no. ST/X508329/1); the US Department of Energy (USDOE) through the Los Alamos National
Laboratory (operated by Triad National Security,
LLC, for the National Nuclear Security Administration
of the USDOE, contract 89233218CNA000001); and the YCAA Prize Postdoctoral Fellowship. This work was also supported by the National Research Foundation of Korea (NRF) grant funded by the Korea government(MSIT) (RS-2024-00449206). We acknowledge support from the Coordenação de Aperfeiçoamento de Pessoal de Nível Superior (CAPES) of Brazil through PROEX grant number 88887.845378/2023-00. We acknowledge financial support from Millenium Nucleus NCN23\_002 (TITANs) and Comité Mixto ESO-Chile.

We thank
the staff at the participating observatories, correlation
centers, and institutions for their enthusiastic support.
This paper makes use of the following ALMA data:
ADS/JAO.ALMA\#2017.1.00841.V and ADS/JAO.ALMA\#2019.1.01797.V.
ALMA is a partnership
of the European Southern Observatory (ESO;
Europe, representing its member states), NSF, and
National Institutes of Natural Sciences of Japan, together
with National Research Council (Canada), Ministry
of Science and Technology (MOST; Taiwan),
Academia Sinica Institute of Astronomy and Astrophysics
(ASIAA; Taiwan), and Korea Astronomy and
Space Science Institute (KASI; Republic of Korea), in
cooperation with the Republic of Chile. The Joint
ALMA Observatory is operated by ESO, Associated
Universities, Inc. (AUI)/NRAO, and the National Astronomical
Observatory of Japan (NAOJ). The NRAO
is a facility of the NSF operated under cooperative agreement
by AUI.
This research used resources of the Oak Ridge Leadership Computing Facility at the Oak Ridge National
Laboratory, which is supported by the Office of Science of the U.S. Department of Energy under contract
No. DE-AC05-00OR22725; the ASTROVIVES FEDER infrastructure, with project code IDIFEDER-2021-086; the computing cluster of Shanghai VLBI correlator supported by the Special Fund 
for Astronomy from the Ministry of Finance in China;  
We also thank the Center for Computational Astrophysics, National Astronomical Observatory of Japan. This work was supported by FAPESP (Fundacao de Amparo a Pesquisa do Estado de Sao Paulo) under grant 2021/01183-8.

APEX is a collaboration between the
Max-Planck-Institut f{\"u}r Radioastronomie (Germany),
ESO, and the Onsala Space Observatory (Sweden). The
SMA is a joint project between the SAO and ASIAA
and is funded by the Smithsonian Institution and the
Academia Sinica. The JCMT is operated by the East
Asian Observatory on behalf of the NAOJ, ASIAA, and
KASI, as well as the Ministry of Finance of China, Chinese
Academy of Sciences, and the National Key Research and Development
Program (No. 2017YFA0402700) of China
and Natural Science Foundation of China grant 11873028.
Additional funding support for the JCMT is provided by the Science
and Technologies Facility Council (UK) and participating
universities in the UK and Canada. 
The LMT is a project operated by the Instituto Nacional
de Astr\'{o}fisica, \'{O}ptica, y Electr\'{o}nica (Mexico) and the
University of Massachusetts at Amherst (USA).
The IRAM 30 m telescope on Pico Veleta, Spain and the NOEMA interferometer on Plateau de Bure,
France are operated by IRAM and supported by CNRS (Centre National de la Recherche Scientifique, France), MPG (Max-Planck-Gesellschaft, Germany), and IGN (Instituto Geográfico Nacional, Spain).
The SMT is operated by the Arizona
Radio Observatory, a part of the Steward Observatory
of the University of Arizona, with financial support of
operations from the State of Arizona and financial support
for instrumentation development from the NSF.
Support for SPT participation in the EHT is provided by the National Science Foundation through award OPP-1852617 
to the University of Chicago. Partial support is also 
provided by the Kavli Institute of Cosmological Physics at the University of Chicago. The SPT hydrogen maser was 
provided on loan from the GLT, courtesy of ASIAA.

This work used the
Extreme Science and Engineering Discovery Environment
(XSEDE), supported by NSF grant ACI-1548562,
and CyVerse, supported by NSF grants DBI-0735191,
DBI-1265383, and DBI-1743442. XSEDE Stampede2 resource
at TACC was allocated through TG-AST170024
and TG-AST080026N. XSEDE JetStream resource at
PTI and TACC was allocated through AST170028.
This research is part of the Frontera computing project at the Texas Advanced 
Computing Center through the Frontera Large-Scale Community Partnerships allocation
AST20023. Frontera is made possible by National Science Foundation award OAC-1818253.
This research was done using services provided by the OSG Consortium~\citep{osg07,osg09}, which is supported by the National Science Foundation award Nos. 2030508 and 1836650.
Additional work used ABACUS2.0, which is part of the eScience center at Southern Denmark University, and the Kultrun Astronomy Hybrid Cluster (projects Conicyt Programa de Astronomia Fondo Quimal QUIMAL170001, Conicyt PIA ACT172033, Fondecyt Iniciacion 11170268, Quimal 220002). 
Simulations were also performed on the SuperMUC cluster at the LRZ in Garching, 
on the LOEWE cluster in CSC in Frankfurt, on the HazelHen cluster at the HLRS in Stuttgart, 
and on the Pi2.0 and Siyuan Mark-I at Shanghai Jiao Tong University.
The computer resources of the Finnish IT Center for Science (CSC) and the Finnish Computing 
Competence Infrastructure (FCCI) project are acknowledged. This
research was enabled in part by support provided
by Compute Ontario (http://computeontario.ca), Calcul
Quebec (http://www.calculquebec.ca), and the Digital Research Alliance of Canada (https://alliancecan.ca/en).

The EHTC has
received generous donations of FPGA chips from Xilinx
Inc., under the Xilinx University Program. The EHTC
has benefited from technology shared under open-source
license by the Collaboration for Astronomy Signal Processing
and Electronics Research (CASPER). The EHT
project is grateful to T4Science and Microsemi for their
assistance with hydrogen masers. This research has
made use of NASA's Astrophysics Data System. We
gratefully acknowledge the support provided by the extended
staff of the ALMA, from the inception of
the ALMA Phasing Project through the observational
campaigns of 2017 and 2018. We would like to thank
A. Deller and W. Brisken for EHT-specific support with
the use of DiFX. We thank Martin Shepherd for the addition of extra features in the Difmap software 
that were used for the CLEAN imaging results presented in this paper.
We acknowledge the significance that
Maunakea, where the SMA and JCMT EHT stations
are located, has for the indigenous Hawaiian people.


\bibliography{references}{}
\bibliographystyle{aasjournal}

\allauthors

\end{document}